\documentclass[10pt]{iopart}
\usepackage{xcolor}
\usepackage[utf8]{inputenc}
\usepackage{amsmath}
\usepackage{amssymb}
\usepackage{graphicx}
\usepackage{gensymb}
\usepackage{graphics}
\usepackage[T1]{fontenc}
\usepackage{soul}
\PassOptionsToPackage{hyphens}{url}\usepackage{hyperref}
\begin{document}

\title[Superconductivity in gated molybednum dichalcogenides]{Competition between phonon mediated superconductivity and carriers localization in field-effect doped molybdenum dichalcogenides}
\author{Giovanni Marini}
\address{$^1$Graphene Labs, Fondazione Istituto Italiano di Tecnologia, Via Morego, I-16163 Genova, Italy}
\ead{giovanni.marini@iit.it}
\vspace{10pt}

\author{Matteo Calandra} 
\address{$^2$Department of Physics, University of Trento, Via Sommarive 14, 38123 Povo, Italy}
\address{$^1$Graphene Labs, Fondazione Istituto Italiano di Tecnologia, Via Morego, I-16163 Genova, Italy}
\address{$^3$Sorbonne Universit\'e, CNRS, Institut des Nanosciences de Paris, UMR7588, F-75252 Paris, France}
\ead{m.calandrabuonaura@unitn.it}
\vspace{10pt}

\begin{abstract}
Superconductivity occurs in electrochemically doped molybdenum dichalcogenides samples thicker than four layers.
While the critical temperature (T$_c$) strongly depends on the field effect geometry (single or double gate) and on the sample (MoS$_2$ or MoSe$_2$), T$_c$ always saturates at high doping. The pairing mechanism  and the complicate dependence of T$_c$ on doping, samples and  field-effect geometry  are  currently not understood. Previous theoretical works assumed homogeneous doping of a single layer and attributed the T$_c$ saturation to a charge density wave instability, however the calculated values of the electron-phonon coupling in the harmonic approximation were one order of magnitude larger than the experimental estimates based on transport data.  
Here, by performing fully relativistic first principles calculations accounting for the sample thickness, the  field-effect  geometry and anharmonicity,  we rule out the occurrence of charge density waves in the experimental doping range  and demonstrate a suppression  of one order of magnitude in the electron-phonon coupling, now in excellent agreement with transport data.  By solving the anisotropic Migdal-Eliashberg equations, we explain the behaviour of T$_c$ in different systems and geometries. From an analysis of mobility data, we propose that the T$_c$ saturation is due to carriers localization and disorder.

\end{abstract}
\maketitle

\ioptwocol
\section{Introduction}

The development of ionic-liquid based electric double layer field effect transistors 
has allowed an extreme versatility in doping low dimensional semiconductors. With this experimental technique, the mobile ions in the electrolyte form, upon gating, 
a narrow spatial charge layer at the interface with the semiconductor 
and induce a charge accumulation in the latter \cite{https://doi.org/10.1002/adfm.200801633} which, in the most recent experiments, can be larger than 10$^{14}~e^-$/cm$^{2}$. 
 Among the plethora of interesting phenomena emerging at such high carrier densities\cite{Ye2010,doi:10.1073/pnas.1018388108,doi:10.1063/1.3493190}, an especially intriguing one is the observation of superconductivity in 
 several multilayers of transition metal dichalcogenides. Superconductivity was first reported in molybdenum disulfide (MoS$_2$) by Ye {\it et al.} in 2012\cite{doi:10.1126/science.1228006} and soon after a similar superconducting state has been observed also in other transition metal dichalcogenides\cite{PMID:26235962,doi:10.1073/pnas.1716781115,Costanzo2018,AliElYumin2019}. 
 
 \begin{figure}[h]
\includegraphics[width=0.9\linewidth]{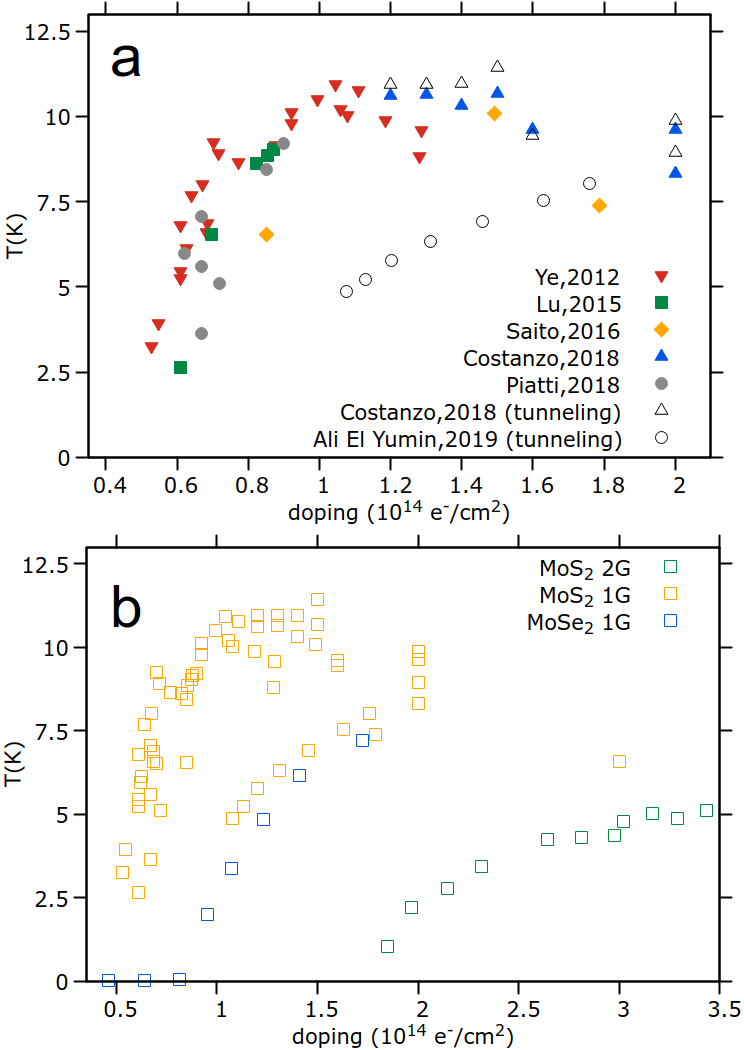}
    \caption{Panel a: measured superconducting critical temperature in MoS$_2$ from transport experiments (filled colored points) and tunneling measurements (empty black points). Experimental data from Refs.\cite{doi:10.1021/acs.nanolett.8b01390,Costanzo2018,doi:10.1126/science.1228006,Saito2016,doi:10.1126/science.aab2277,AliElYumin2019}. Panel b: superconducting critical temperature measured in single-side gated MoS$_2$ (yellow squares), double-side gated MoS$_2$ (green squares, data from Ref.\cite{Zheliuk2019}) and single-side gated MoSe$_2$ (blue squares, data from Ref.\cite{PMID:26235962}).}\label{fig1}
\end{figure}

 The main experimental findings regarding superconductivity in gated molybdenum dichalcogenides are illustrated in Fig.  \ref{fig1}. All measurements report a superconducting region in MoS$_2$ in a single gate geometry for thickness larger than four layers. However, different measurements show substantially different results. In Refs.  
 \cite{doi:10.1126/science.1228006,doi:10.1073/pnas.1716781115} a superconducting region emerges at  induced charges (doping) of $n_e\approx0.5\times10^{14}$ e$^- /{\rm cm}^2$ with T$_c$ rapidly increasing up to $10$ K at $n_e \approx 0.9\times10^{14}$ e$^- /{\rm cm}^2$.
 At larger doping, Refs. \cite{doi:10.1126/science.1228006,Costanzo2018,doi:10.1073/pnas.1716781115,Saito2016} show a flattening (saturation) in the T$_c$ versus doping curve that has been interpreted as a {\it dome} albeit the experimental data are substantially scattered (variations
 in T$_c$ of the order of $3$ K occurs at $n_e\approx 1.3\times10^{14}$ e$^- /{\rm cm}^2$) and are also consistent with a doping independent T$_c$. The measurements of Ref. \cite{AliElYumin2019} step aside from the others as
 superconductivity increases with very weak signs of saturation at large doping ($n_e > 1.1\times10^{14}$ e$^- /{\rm cm}^2$). The superconducting T$_c$ measured in this last experiment are somewhat lower than those in Refs. 
 \cite{doi:10.1126/science.1228006,Costanzo2018,doi:10.1073/pnas.1716781115,Saito2016}. Finally, it is important to mention that in MoS$_2$ in a single gate geometry the superconducting transition also depends on the thickness of the sample: namely, monolayer samples show very low  superconducting critical temperatures (T$_c\approx 2$ K or no superconductivity \cite{Yecomm}) with respect to their few-layer counterparts\cite{Costanzo2016} and with superconductivity, when occurring, confined in the very  low doping region. 
 
 The experimental findings change substantially if a double layer geometry is considered.  The case of a MoS$_2$ bilayer in a double gate geometry  is particularly interesting as the equal charging of the gates does not break the mirror symmetry with respect to the plane (as it happens, on the contrary, in the single gate geometry). In this case, it has been shown in Ref.\cite{Zheliuk2019} that superconductivity only occurs at doping larger than $n_e=1.8\times 10^{14}$ and T$_c$ increases smoothly up to $\approx 5$K at $n_e=3.45\times 10^{14}$(see Fig. \ref{fig1} bottom panel). 
 
 Finally, in the case of multilayer molybdenum diselenide (MoSe$_2$) (thickness 20-100 nm\cite{PMID:26235962}) in a single gate geometry,  it has been shown that superconductivity emerges at  $n_e\approx 1.0\times10^{14}$ e$^- /{\rm cm}^2$ and increases up to $\approx 7.5$ K at the highest achievable doping of $\approx 1.7\times10^{14}$ e$^- /{\rm cm}^2$.\cite{PMID:26235962}

All these experimental observations call for a comprehensive theoretical explanation of superconductivity in gated transition metal dichalcogenides, that has, however, remained elusive due to the complex dependence of their superconducting properties on many different factors. Previous theoretical works \cite{PhysRevB.87.241408,PhysRevB.90.245105,Fu2017} investigated superconductivity
treating the charging of the flake in the framework of a uniform background doping.
The large electric field and the resulting highly inhomogeneous doping charge distribution along the direction perpendicular to the flake were completely neglected.
Moreover, most of these works investigate the case of a single layer 
\cite{PhysRevB.87.241408,PhysRevB.90.245105} and compared their results with measurements on thicker flakes assuming the thickness  to be irrelevant.
In Ref. \cite{Fu2017} the inadequacy of the uniform 
doping charge distribution was partially demonstrated by showing that
simulating a single layer MoS$_2$ in an homogeneous doping background (i.e. neglecting the role of the electric field) does  result in a superconducting state, but simulating a bilayer does not, in striking disagreement with experimental data.
Furthermore, calculations on a single layer and a uniform doping \cite{PhysRevB.87.241408,PhysRevB.90.245105,Fu2017} showed an anomalously large electron-phonon interaction reaching values as large as $\lambda\approx 8$. These values are more than one order of magnitude larger than the experimental estimates based on transport data in Ref. \cite{AliElYumin2019}.
In order to find
T$_c$ comparable with experiments the authors were forced to use unphysical values of $\mu^*$ ( $\mu^*=0.25$ or more). Furthermore, the apparent superconducting dome as a function of doping predicted in Ref.\cite{PhysRevB.90.245105} is a spurious result of the Allen-Dynes parametrization at high $\lambda$ values\cite{PhysRevB.12.905}: we verified that the correct description using the Eliashberg equation results in an ever increasing superconducting T$_c$. Moreover this large electron-phonon interaction was found to result in a  charge density wave (CDW) at $n_e=1.17\times 10^{14} e^-/{\rm cm}^2$. The authors of Ref. \cite{Roesner2017} attributed the flattening of T$_c$ at high doping  to this emerging CDW \cite{PhysRevB.90.245105}. However these results were obtained by neglecting the presence of the electric field and, most important, by neglecting anharmonic effects that should become crucial close to the CDW instability.

It is worthwhile to underline that all these works neglected relativistic effects that are known to be large in these system and are supposed to be even larger in the presence of the electric field breaking the inversion symmetry of the flake.

An alternative model description of the superconducting state has been attempted invoking an unconventional $s\pm$ pairing originating from Coulomb interaction\cite{PhysRevB.88.054515}. However this explanation relies on the assumption of a larger inter-valley than intra-valley electron-electron repulsion and assumes the presence of an high-$\kappa$ dielectric substrate. The first assumption seems to be in contrast with first principles estimation of the Coulomb interaction \cite{PhysRevB.94.134504}, the second is not backed up by experiments as superconductivity occurs even in the absence of an high-$\kappa$ dielectric substrate. 

After having reviewed these results in literature, we can state that a clear understanding of the mechanism leading to superconductivity in samples doped in ionic-liquid based field-effect transistors (FET)  is missing. In particular, as most of previous work only addressed one compound and neglected the field effect geometry, a unified picture of the intricate dependence of T$_c$ on the compound and on the FET geometry is clearly lacking.

In this work, by using first principles electronic structure calculations, we investigate the occurrence of superconductivity in MoS$_2$ and MoSe$_2$ multilayers in single gate geometry and in bilayer MoS$_2$ in double gate geometry. Our calculations explicitly include (i) the FET geometry and the consequent charge inhomogeneity in the direction perpendicular to the layers and to the gate,
(ii) anharmonic effects, (iii) the thickness of the sample, and (iv) relativistic effects. By solving the anisotropic multi-band Eliashberg equations we investigate superconducting properties and compare our findings with  experimental data. Finally, we give a comprehensive explanation of the mechanism leading to superconductivity in gated transition metal dichalcogenides and discuss the possible role of the disorder and charge inhomogeineities in determining T$_c$ saturation.
Our work provides the first complete explanation of superconductivity in this class of materials and, beside demonstrating  that superconductivity is phonon mediated, it shows that the measured T$_c$ are due to the cooperation of several phenomena: thickness of the sample, strong electric-fields, anharmonicity and charge disorder. 

The paper is structured as follows: In Sec. \ref{sec2} we give the technical details of our first principles calculations, In Sec. \ref{sec3} we discuss the thickness dependence of the electronic and vibrational properties of gated molybdenum dichalcogenides, in Sec. \ref{sec4} we study the field-effect doping dependence of the electronic properties, while Sec. \ref{sec5} is dedicated to the analysis of the vibrational and electron-phonon coupling properties. Finally, in Sec.\ref{sec6} we discuss the superconducting properties of molybdenum dichalcogenides, in Sec. \ref{sec:disorder} we discuss the role of charge inhomogeneities in determining T$_c$ saturation and, finally, in Sec. \ref{sec7} we give our conclusions.

\section{Technical Details}\label{sec2}

\begin{table*}
    \centering
    \begin{tabular}{|c|c|c|c|c|c|}
    \hline
 Material & $e^-$/cell & $10^{14}~e^-/$cm$^2$ & in-plane parameter & interlayer distance & Mo-chalcogen distance\\
 \hline
 \hline
    1G MoS$_2$ & 0 & 0 & 3.16 & 6.25 & 1.57\\
    \hline
      1G MoS$_2$ & 0.1 & 1.15 & 3.16 & 6.24 & 1.57\\
    \hline
      2G MoS$_2$ & 0 & 0 & 3.16 & 6.25 & 1.57\\
        \hline
      2G MoS$_2$ & 0.3 & 3.36 & 3.20 & 7.47 & 1.58\\
    \hline
    1G MoSe$_2$ & 0 & 0 & 3.28 & 6.55 & 1.68\\
    \hline
    1G MoSe$_2$ & 0.15 & 1.59 & 3.28 & 6.54 & 1.68\\
    \hline
    \end{tabular}
    \caption{Structural parameters used in the calculations for bilayer single-side gated MoS$_2$, double-side gated MoS$_2$ and single-side gated MoSe$_2$ as a function of doping.}
    \label{tab1}
\end{table*}

We perform first-principles calculations 
using the {\sc Quantum ESPRESSO} package\cite{QE,QE2}. We use full-relativistic Optimized Norm-Conserving Vanderbilt (ONCV) pseudopotentials\cite{doi:10.1021/acs.jctc.6b00114}, including semi-core states. We choose  a kinetic energy cutoff of 80~Ry.  We adopt the generalized gradient approximation (GGA) in the Perdew, Burke and Ernzerhof (PBE)\cite{PBE} formulation for the exchange-correlation potential and include long range dipolar van-der-Waals corrections within the approach of Ref.\cite{https://doi.org/10.1002/jcc.20495} to improve the description of interlayer interaction. We use a {16$\times$16$\times$1}  Monkhorst-Pack wave-vector grid\cite{MP}  for the electron-momenta  Brillouin zone (BZ) integration in the self-consistent charge density calculation and a Gaussian smearing of 0.0075~Ry. Spin-orbit copuling is included in the calculations. 

We simulate the field-effect setup (both the single-side and double-side gating) according to the methodology presented in Ref.\cite{PhysRevB.96.075448}. We employ a constant potential energy barrier of 2.5 Ry and thickness $\approx$ 0.03~\AA~  to mimic the effect of the electrostatic barrier due to the atoms/molecules of the gate dielectric. In the double-side gating configuration we assume that both charged plates possess the same amount of charge. A vacuum larger than 20 \AA~is used in all calculations to avoid that the boundaries of the physical region in the approach of Ref.\cite{PhysRevB.96.075448} are too close to the material. 

The structural parameters of the different compounds are reported in Tab.\ref{tab1}. For the case of single-side gated compounds, we fix the in plane lattice parameter to the experimental values\cite{PMID:26235962} (3.16 \AA~ for MoS$_2$, 3.28 \AA~ for MoSe$_2$)  and we let the system relax in the out of plane direction. In the absence of FET doping, we find an interlayer spacing of 6.25 \AA~for bilayer MoS$_2$ and 6.55 \AA~ for bilayer MoSe$_2$. The interlayer distance remains almost constant in the single-side gating setup (less than 0.4$\%$ variation between 0 and 1.725 $\times 10^{14}$ e$^-$/cm$^{2}$ for MoS$_2$), while it sensibly increases in the case of double-side gating (from 6.25 at 0 to 6.35 \AA~ at $n_e=1.725 \times 10^{14}$ e$^-$/cm$^{2}$ for MoS$_2$, a 1.6\% increase). For the case of double-side gated bilayer MoS$_2$ the effects of the induced charge on the in-plane lattice parameter is included in the calculation when not differently specified. In particular, we evaluate the in-plane lattice parameter  in a variable-cell simulation as a function of the doping. The expanded in-plane lattice parameter $a_{n}$ is then calculated as 

\begin{equation} a_{n} = a_{exp} * a^{rel}_{n}/a^{rel}_{GS}\end{equation}

where $a^{rel}_{n}$ is the in-plane lattice parameter calculated in the variable-cell relaxation at doping $n$, $a^{rel}_{GS}$ is the equilibrium DFT-PBE lattice parameter and $a_{exp}$ is the experimental lattice parameter.

\begin{figure}[h]
\includegraphics[width=0.9\linewidth]{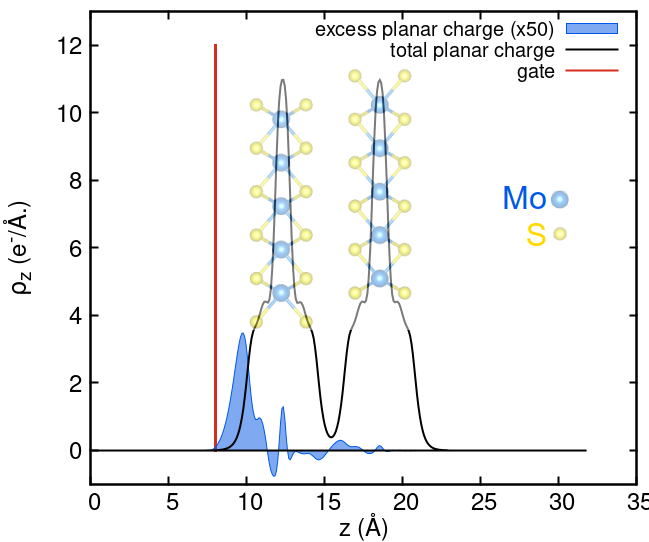}
    \caption{Averaged in-plane charge-density as a function of the out of plane coordinate (black line). Filled blue curve represents the excess charge with respect to the neutral system.}\label{fig2}
\end{figure}

The setup for single-side gated MoS$_2$ is shown in Fig. \ref{fig2}.

The harmonic phonon frequencies are evaluated within density-functional perturbation theory\cite{RevModPhys.73.515} on a 8$\times$8$\times$1 phonon momentum grid ($\mathbf{q}$-grid) and 16$\times$16$\times$1 electron-momentum grid ($\mathbf{k}$-grid). The anharmonic phonon frequencies were calculated within the stochastic self-consistent harmonic approximation (SSCHA), according to the procedure described in Ref.\cite{Monacelli_2021}. The SSCHA accounts for  quantum, thermal and anharmonic effects of nuclei within the Born-Oppenheimer approximation\cite{Monacelli_2021} and requires calculations of forces in a supercell with ionic coordinates displaced from equilibrium following a temperature dependent Gaussian probability. We assume T=0 in the SSCHA calculations. Within this approach, anharmonic phonon frequencies can be computed by diagonalization of the Hessian of the minimized SSCHA free energy.  
Supercell calculations were performed on a 4$\times$4$\times$1 supercell (96 atoms) and a 4$\times$4$\times$1 momentum grids for Brillouin zone integrals in the supercell. In order to include the effect of anharmonicity in the calculation of superconducting properties, we use anharmonic phonon frequencies together with the electron-phonon matrix elements evaluated at the harmonic level, to calculate $\lambda$ and the isotropic Eliashberg function $\alpha^2F(\omega)$. 

The correct evaluation of the electron-phonon coupling properties of doped few-layer MoS$_2$ requires a precise knowledge of electron-phonon matrix elements for very dense electron and phonon momentum grids. Since the direct calculation of electron-phonon matrix elements over a ultradense $\mathbf{q}$- and $\mathbf{k}$-point grids is very time consuming  in linear response, we perform a Wannier interpolation of the electron-phonon coupling as described in  Ref.\cite{PhysRevB.82.165111}.
We used the Wannier90 \cite{MOSTOFI2008685} code to obtain the Bloch to Wannier transformation. We use as starting guess of the Maximally Localized Wannier Function procedure three $p$-like orbitals at every chalcogen site and 5 $d$-like orbitals at any molybdenum site. Within this approach, the electron-phonon matrix elements are first calculated on a coarse 8$\times$8$\times$1 $\mathbf{q}$ grid and 16$\times$16$\times$1 $\mathbf{k}$-grid, and then Wannier interpolated to 64$\times$64$\times$1 $\mathbf{q}$- and $\mathbf{k}$- grids in order to evaluate the electron-phonon coupling parameter $\lambda$ and the isotropic Eliashberg function $\alpha^2F(\omega)$. We employ a Gaussian smearing of 0.001~Ry for $\mathbf{k}$- and $\mathbf{q}$- summations in $\lambda$.

\begin{figure}[h]
\includegraphics[width=1\linewidth]{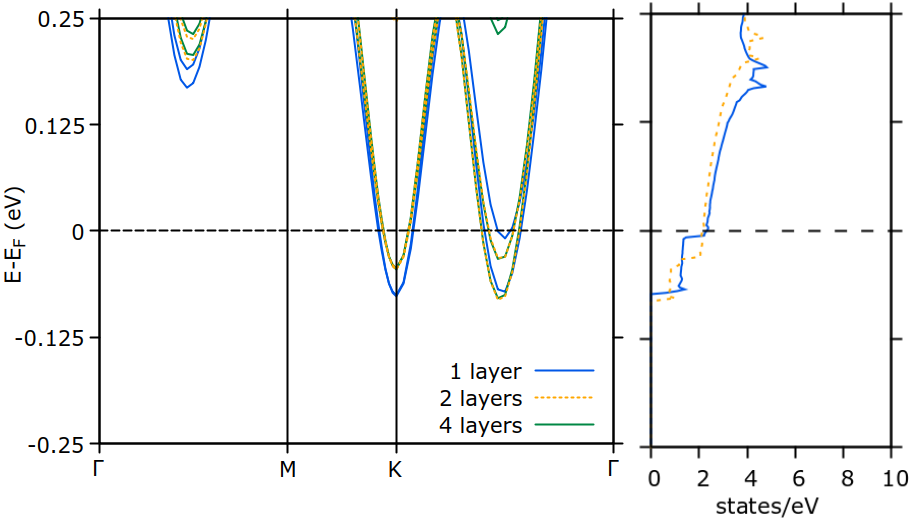}
    \caption{Evolution of the low-energy band-structure and density of states with the number of layers  for single-side gated MoS$_2$ at 1.15 $\times 10^{14}$ e$^-$/cm$^{2}$.}\label{fig3}
\end{figure}

The superconducting gap is evaluated by solving in the Wannier basis the Migdal-Eliashberg equations \cite{AllenMitrovic_TheoryofSuperconductingTc_1983} over the imaginary frequency axis and then by performing analytic continuation to the real axis using N-point Pad\'e approximants\cite{Vidberg1977}. We sample the Fermi surface using 2048 randomly generated $\mathbf{k}$- and $\mathbf{k+q}$- points. At least 128 Matsubara frequencies were necessary to converge the solution for the gap on the imaginary axis. 
The superconducting critical temperature is then evaluated by determining the temperature where the superconducting gap becomes zero. We usa a Morel-Anderson pseudopotential\cite{PhysRev.125.1263} $\mu^*$ to parametrize the Coulomb repulsion in the superconducting state. A $\mu^* = 0.1$ is used when not otherwise specified. The spin-orbit coupling is not included in the solution of Migdal-Eliashberg equations. It was however verified that the inclusion of spin-orbit coupling does not alter the calculated electron-phonon coupling value appreciably.

\begin{figure}[h]
\includegraphics[width=0.9\linewidth]{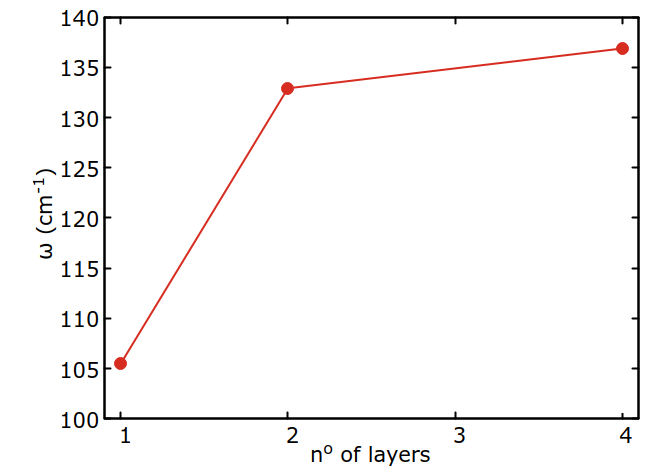}
    \caption{Lowest phonon frequency at the M point as a function of the number of layers for single-side gated MoS$_2$ at  1.15 $\times 10^{14}$ e$^-$/cm$^{2}$.}\label{fig4}
\end{figure}

\section{Convergence of electronic and vibrational properties in the field-effect setup as a function of thickness}\label{sec3}

The first necessary step is to understand how the electronic and vibrational properties of gated MoS$_2$ are affected by the presence of the charging electric field in field-effect geometry and how the properties depend on the number of layers. Our aim is to understand what is the minimal thickness of the multilayer that is representative of the  thick samples (thicker than 4 layers) used in experiments where superconductivity has been detected.

We first consider a MoS$_2$ bilayer in FET geometry with a single gate on one side. We perform complete structural optimization and calculate the planar average
\begin{eqnarray}
\rho_z =\frac{e}{\Omega N_k}\sum_{{\bf k} n} \int_{\Omega}  |\psi_{{\bf k} n}({\bf r})|^2\,dA
\end{eqnarray}
where $\Omega$ is the surface of the 2D unit cell and $\psi_{{\bf k} n}({\bf r})$ are the Kohn-Sham eigenfunctions. The result of the calculation is the black line show in Fig. \ref{fig2}. However, as this quantity is dominated by the contribution of semicore
electrons (and non-valence electrons in general), we also calculate the excess charge with respect to the neutral MoS$_2$ bilayer, namely 
$\rho_z({\rm FET~doped})-\rho_z({\rm undoped})$. As it can be seen in Fig. \ref{fig2} (filled blue curve), the excess charge distributes non uniformly and is mostly localized in the first layer of S atoms close to the gate. This clearly invalidates the treatment of a uniform doping even for a single layer.

\begin{figure}[h]
\includegraphics[width=1\linewidth]{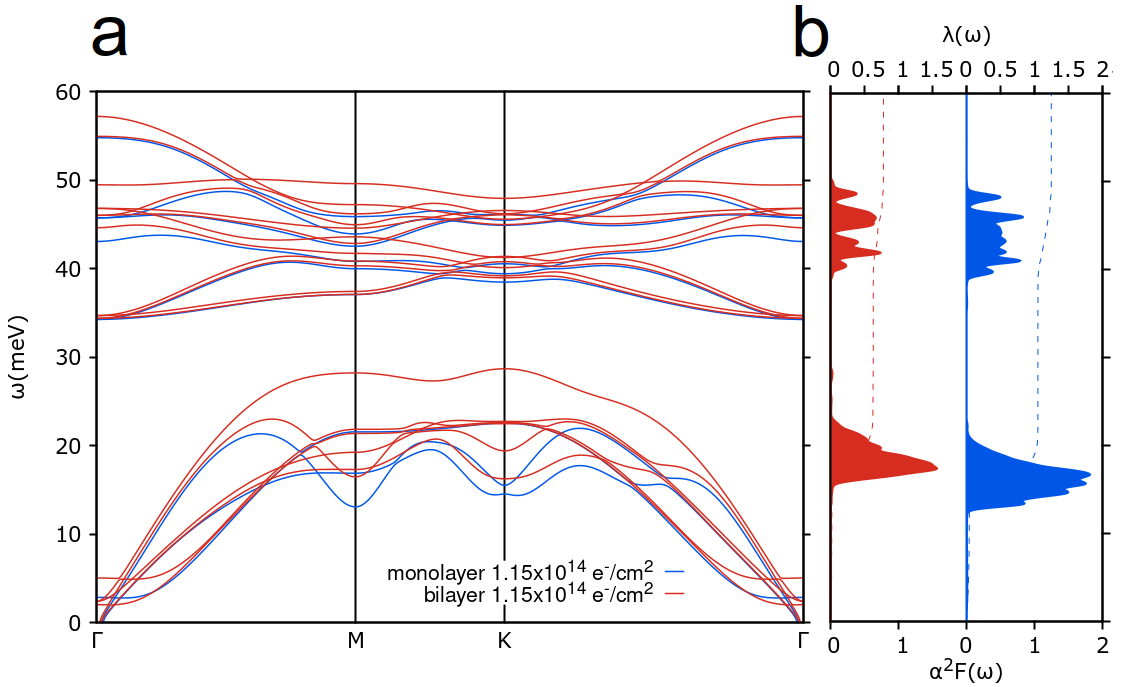}
    \caption{Phonon dispersion for monolayer (blue) and bilayer (red) MoS$_2$ in FET geometry at 1.15 $\times 10^{14}$ e$^-$/cm$^{2}$ (panel a) and corresponding Eliashberg function $\alpha^2F(\omega)$ (filled curves) and $\lambda(\omega)$ (dashed lines) (panel b).}\label{fig5}
\end{figure}

\begin{figure*}[htpb!]
\includegraphics[width=1\linewidth]{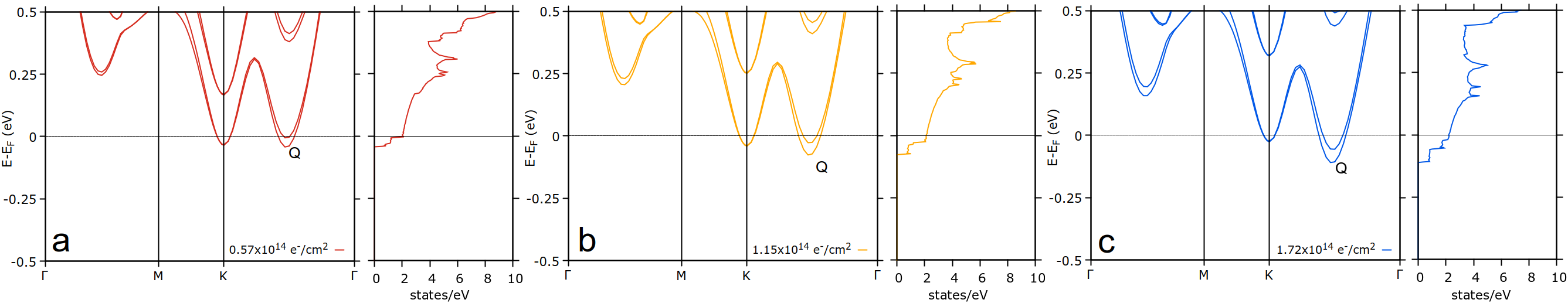}
    \caption{Electronic structure as a function of doping in single-side gated MoS$_2$. The doping level are $n_e=$0.57 $\times 10^{14}$ e$^-$/cm$^{2}$, $n_e=$1.15 $\times 10^{14}$ e$^-$/cm$^{2}$ and n$_e=$1.72 $\times 10^{14}$ e$^-$/cm$^{2}$ in panels (a), (b) and (c), respectively}\label{fig6}
\end{figure*}

We then perform DFT simulations of single-side gated MoS$_2$ employing an increasingly higher number of layers, starting from one and up to four. In Fig. \ref{fig3} we show the evolution of the bottom of the conduction band as a function of the number of layers in the single gate geometry. As it can be seen, the $2$ and $4$ layers electronic structures are practically indistinguishable. Furthermore we also check the dependence of the lowest phonon frequency at the $M$ point of the hexagonal lattice that crucially depends on the excess charge density distribution as a function of the number of layers (see Fig.  \ref{fig4}) at a doping of $n_e=$1.15 $\times 10^{14}$ e$^-$/cm$^{2}$. As it can be seen also the value of the phonon frequency is substantially different from one to two layers, with the single layer case being substantially softer. On the contrary the bilayer is practically identical to the four layer case.

\begin{figure}[htpb!]
\centering
\includegraphics[width=0.8\linewidth]{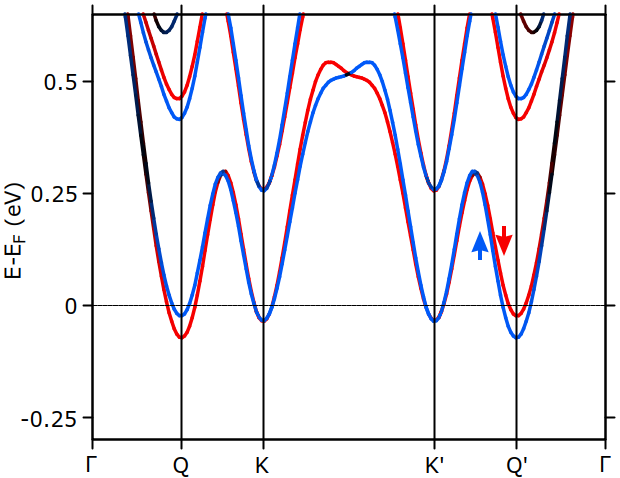}
    \caption{Electronic structure for single-side gated MoS$_2$ at $n_e=$1.15 $\times 10^{14}$ e$^-$/cm$^{2}$ along the diagonal direction of the BZ, projected over the up (red) and down (blue) out-of-plane spin component. Here, black color indicates the absence of any out-of-plane spin character.}\label{fig7}
\end{figure}

In Fig. \ref{fig5} a, we compare the vibrational spectrum of single-side gated monolayer and bilayer MoS$_2$ in the FET geometry at doping concentration  $n_e = 1.15\times$ 10$^{14}~e^-$/cm$^{2}$. This is approximately the value where the superconducting T$_c$ maximum is reached in the experiments. The plot highlights a general tendency of the monolayer phonon frequencies to be lower with respect to the bilayer. This, in combination with an enhanced electron-phonon coupling strength (corresponding to $\approx$ 22.5 \% increase in the first momentum of $\alpha^2$F) results in a generally higher value for the Eliashberg function in the monolayer compared to the bilayer (see Fig. \ref{fig5} b) and a consequent higher value of the electron-phonon coupling parameter $\lambda$ (see sec. \ref{sec:SGMoS2} for definition of Eliasberg function $\alpha^2F(\omega)$ and of $\lambda$) , despite having a practically equal density of states at the Fermi level (see Fig. \ref{fig3}).

\begin{figure*}[htpb!]
\includegraphics[width=1\linewidth]{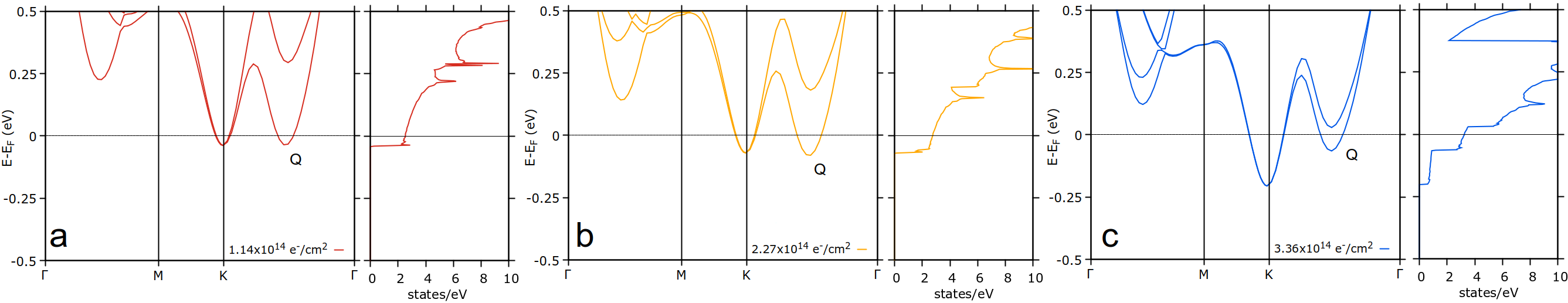}
    \caption{Electronic structure as a function of doping in double-side gated MoS$_2$. The doping levels are 
    $n_e=$1.15 $\times 10^{14}$ e$^-$/cm$^{2}$, 
    $n_e=$2.3 $\times 10^{14}$ e$^-$/cm$^{2}$,
    $n_e=$3.45 $\times 10^{14}$ e$^-$/cm$^{2}$ in panels (a), (b) and (c), respectively}
   \label{fig8}
\end{figure*}

\begin{figure*}[htpb!]
\includegraphics[width=1\linewidth]{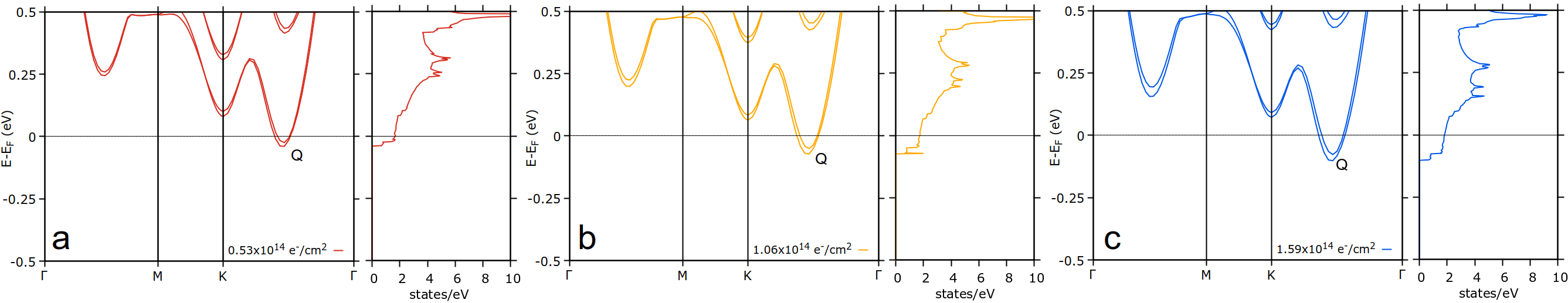}
    \caption{Electronic structure as a function of doping in single-side gated MoSe$_2$. The doping levels are 
    $n_e=$0.53 $\times 10^{14}$ e$^-$/cm$^{2}$,
    $n_e=$1.06 $\times 10^{14}$ e$^-$/cm$^{2}$,
    $n_e=$1.59 $\times 10^{14}$ e$^-$/cm$^{2}$, in panels 
    (a), (b) and (c), respectively.
    }\label{fig9}
\end{figure*}

\begin{figure}[htpb!]
\includegraphics[width=0.8\linewidth]{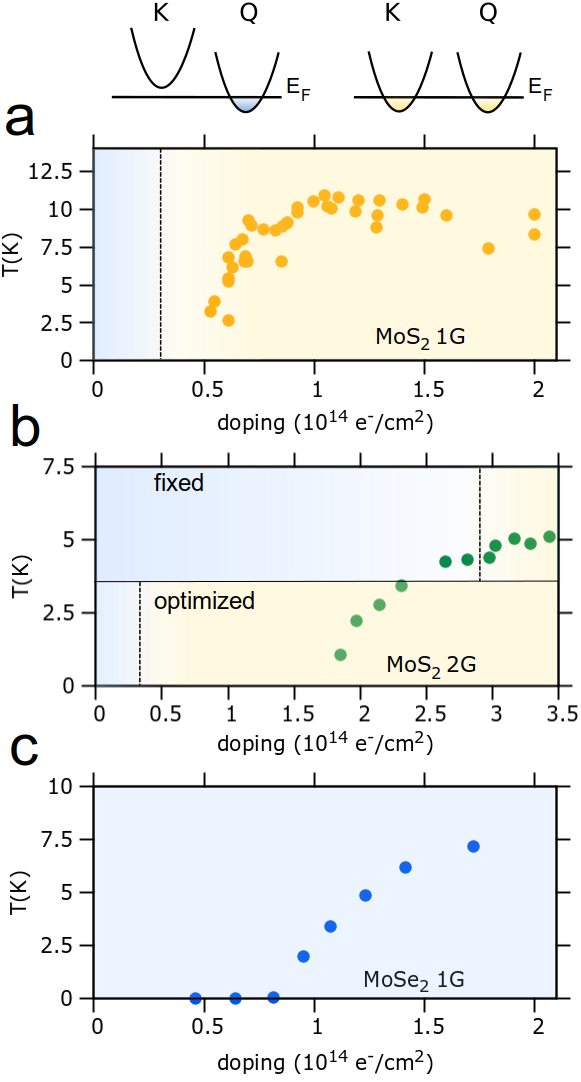}
    \caption{Experimentally measured superconducting critical temperature as a function of doping data for single-side gated MoS$_2$ (panel a), double-side gated MoS$_2$ (panel b), single-side gated MoSe$_2$ (panel c). Experimental data taken from Refs.\cite{Costanzo2018,doi:10.1126/science.1228006,PMID:26235962,Saito2016,doi:10.1126/science.aab2277}. Light blue and yellow backgrounds indicate the areas where only the minimum at $\bf{Q}$ (light blue) and both $\bf{K}$ and $\bf{Q}$ minima are populated (yellow).}\label{fig10}
\end{figure}

Conversely, it is known from experiments that under a certain thickness, and especially for a monolayer, a degradation of superconducting properties occurs\cite{Costanzo2016,Yecomm}. 
As our simulations predict that the electron-phonon coupling in the monolayer is generally higher than in the bilayer, we can exclude that the degradation of superconducting properties observed in monolayer is related to a lower electron-phonon coupling: a different explanation should be invoked to explain this observation, see $e.g.$ the localization of carriers discussed in Sec. \ref{sec:disorder} of this work and also discussed  in Ref.\cite{doi:10.1073/pnas.1716781115} for monolayer WS$_2$.

As we find that monolayer and bilayer show sensibly different electronic and vibrational properties, while the latter has very similar properties to the four-layer
system, we establish that the minimal thickness 
that is representative of thicker flakes is a gated bilayer. We will then consider this system as representative of single gated thicker flakes from now on.

\section{Electronic structure as a function of doping in field-effect setup}\label{sec4}

We now switch to the discussion of the electronic properties of gated MoS$_2$ and MoSe$_2$. In Fig. \ref{fig6}, we plot the evolution of the Kohn-Sham energy bands and the corresponding density of states as a function of doping for  single-side gated MoS$_2$. We observe that both conduction band minima at $\bf{K}$ (twofold degenerate in the BZ) and $\bf{Q}$ (six times degenerate) are already occupied at $n_e = 0.57\times$ 10$^{14}~e^-$/cm$^{2}$ and both minima always remain occupied up to the highest considered doping concentration of $n_e = 1.72\times$ 10$^{14}~e^-$/cm$^{2}$. The inclusion of spin-orbit coupling in the calculation induces a large splitting in correspondence to the $\bf{Q}$-point and a minor splitting at the $\bf{K}$-point. The spin-orbit coupling splitting is only possible due to the inversion symmetry breaking caused by the presence of the gate. We find that the $\bf{Q}$ minimum is the first to be populated in single-side gated bilayer MoS$_2$, while the $\bf{K}$ minimum is populated at $n_e\approx$ $0.3\times$ 10$^{14}~e^-$/cm$^{2}$. This result is at variance with a previous first principles investigation on single-side gated MoS$_2$, where the $\bf{K}$ minimum was populated first also in the bilayer case\cite{PhysRevB.91.155436}: the difference is traceable to the usage of the DFT-PBE relaxed in-plane lattice parameter ($\approx$ $3.19$~\AA) instead of the experimental one (3.16~\AA). While the precise choice of the lattice parameter considered in the simulation can in principle affect the superconducting properties, no qualitative differences are expected in the doping region where the superconductivity is observed experimentally both using the DFT-PBE and the experimental parameter\cite{PhysRevB.91.155436}, as both $\mathbf{K}$- and $\mathbf{Q}$- minima are filled\cite{Piatti_2019}. 

In Fig. \ref{fig7}, we report the spin-orbited conduction band structure for single-side gated MoS$_2$ along the diagonal direction of the BZ at at $n_e=$1.15 $\times 10^{14}$ e$^-$/cm$^{2}$, projected onto the out-of-plane spin component. We find that the spin direction is essentially out-of-plane both at $\mathbf{K}$ and $\mathbf{Q}$ minima, with splitted electronic states possessing a coherent up or down character. We report a spin-splitting of $\approx$ 50 meV for the $\bf{Q}/\bf{Q}'$ minima at the Fermi level and of $\approx$ 1.5 meV for the $\bf{K}/\bf{K}'$ minima. The evolution of the electronic band structure as a function of doping with the simultaneous population of $\bf{K}$ and $\bf{Q}$ minima in the conduction band demonstrates that superconductivity in single-side gated MoS$_2$ cannot be simply interpreted in terms of a $\mathbf{K}$-$\mathbf{K}'$ inter-valley pairing. In particular, this implies that a more complex description simultaneously including the population of both $\bf{K}$ and $\bf{Q}$ minima is required to describe the Ising protection of the superconducting state observed experimentally\cite{doi:10.1073/pnas.1716781115,Saito2016,doi:10.1126/science.aab2277}.

A completely different scenario is observed for double-side gated MoS$_2$, which electronic structure is reported in Fig.  \ref{fig8}. Both minima are occupied up to high doping values ($n_e\approx$ 3 $\times$ 10$^{14}~e^-$/cm$^{2}$), and no spin-orbit induced splitting can be found  anywhere in the electronic band structure, as the inversion symmetry is not broken in the presence of double gating (as long as the same amount of charge is present on both gates). Furthermore we note that the minimum at $\bf{K}$ is doubly degenerate (four times degenerate, considering spin degeneracy) in double-side gated MoS$_2$  due to the presence of inversion symmetry, at variance with the single-side gated case. We stress that in experiments the inversion symmetry could be broken also in double-side gated MoS$_2$, due to a finite charge difference between the two gates. However, we do not expect the magnitude of this effect to be large.  In the case of bilayer double-side gated MoS$_2$ we also include the effect of the in-plane structural optimization  induced by doping since the charging is substantial.

The case of single-side gating of MoSe$_2$ is again different. The electronic band structure as a function of doping  is reported in Fig. \ref{fig9}. We observe that  the minimum at $\bf{Q}$ is the only one to be populated in the doping range considered here ($i.e.$ up to $n_e=1.59 \times 10^{14}$e$^-$/cm$^{2}$). Furthermore, a large spin-orbit coupling splitting is observed in correspondence of both $\bf{K}$ and $\bf{Q}$ minima, at variance with the case of single-side gated MoS$_2$. 

Overall, the three considered systems display remarkably different electronic properties. In Fig. \ref{fig10} the experimentally measured superconducting critical temperature for the three cases are reported together with a schematic representation of the occupied minima as a function of doping. We observe that superconductivity in single-gated MoS$_2$ flakes occurs in  the region where both minima are occupied (panel a). 

In Fig. \ref{fig10} b, we include the effect of the in-plane lattice parameter optimization due to the induced charge in evaluating the  occupation of the conduction band minima. This effect  is especially relevant in this case, as a lattice expansion as small as 0.3\% completely changes the critical doping necessary to pass from a single ($\mathbf{Q}$) to a double ($\mathbf{Q}$ and $\mathbf{K}$) conduction-band valley occupation.  This has, of course, an important impact on the predicted superconducting critical temperature, as we will discuss in Sec.\ref{sec6}.

In the case of single-gated MoSe$_2$ superconductivity occurs when only the $\mathbf{Q}$ valley is occupied. Thus,  the occupation of the minimum at $\mathbf{Q}$ is crucial for the occurrence of superconductivity. The reason is the in-plane character of the states forming the conduction band minimum at $\bf{Q}$ that leads to a large electron-phonon coupling, as already noted in Ref. \cite{PhysRevB.87.241408}. 

Finally, we comment on the evolution of the density of states at the Fermi level, $N(0)$, as a function of doping, as it is especially relevant for the electron-phonon coupling properties. In the single-side gated systems, we observe a weak increase as of $N(0)$ as a function of increasing doping in the investigated doping range (8\% for MoS$_2$ from 0.57. to 1.72 $\times$ 10$^{14}~e^-$/cm$^{2}$, 9\% for MoSe$_2$ from 0.53 to 1.59 $\times$ 10$^{14}~e^-$/cm$^{2}$). Thus, we can immediately conclude that the density of states variation is not at the origin of the T$_c$ versus doping behaviour in single gated samples.
On the contrary, a significant 30\% increase is instead calculated for double-side gated MoS$_2$ going from $n_e=1.14$ $\times$ 10$^{14}~e^-$/cm$^{2}$ to $3.36$  $\times$ 10$^{14}~e^-$/cm$^{2}$.

\section{Vibrational properties and electron-phonon interaction}\label{sec5}

Having discussed the dependence of the electronic properties on FET doping for gated MoS$_2$ and MoSe$_2$, we now consider how vibrational properties and electron-phonon interaction  behave as a function of doping. 

\subsection{Single-side gated MoS$_2$\label{sec:SGMoS2}}

\begin{figure}[htpb!]
\includegraphics[width=1\linewidth]{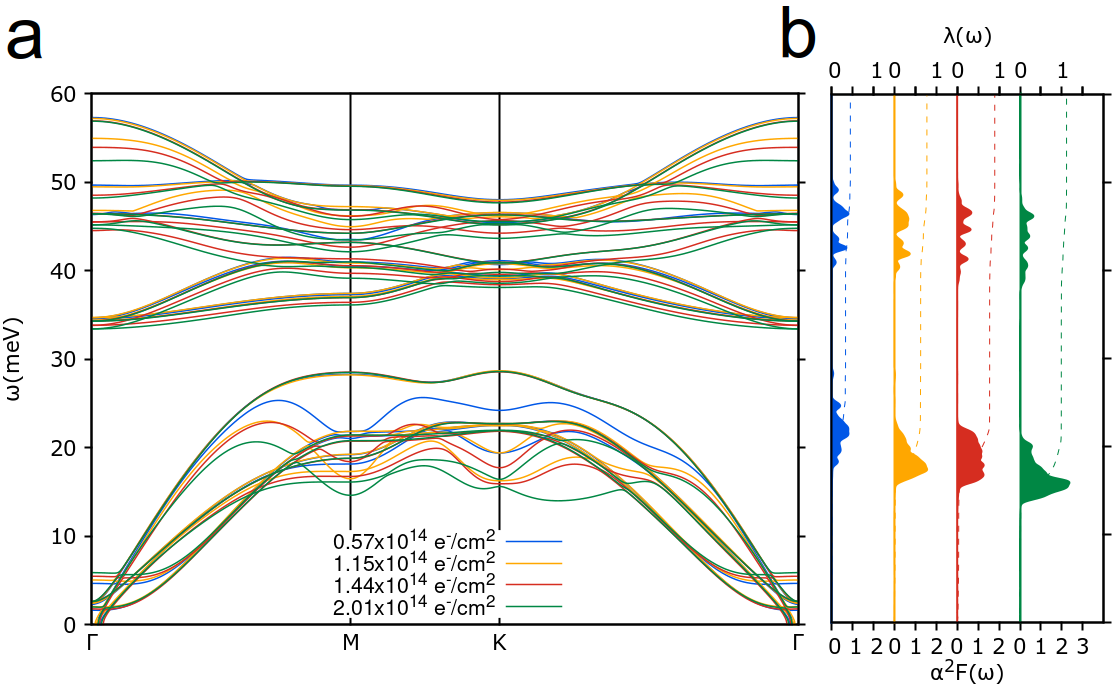}
    \caption{Panel a: anharmonic phonon dispersion as a function of electron doping for single-side gated MoS$_2$. Panel b: corresponding Eliashberg function $\alpha^2F(\omega)$ (filled curves) and $\lambda({\omega})$ (dashed lines). Here, the anharmonic corrections to phonon frequencies is included in the doping range where they become relevant ($n_e$ $\geq$ 1.44 $\times$ 10$^{14}~e^-$/cm$^{2}$).}\label{fig11}
\end{figure}

In Fig. \ref{fig11} a we plot the anharmonic phonon dispersion  as a function of increasing doping for single-side gated MoS$_2$.  Softenings of the lowest-energy phonon band appear with increasing doping, specifically around the M-point and in the K-$\Gamma$ direction of the BZ. These softenings  are related to the inter-band nesting developing between $\bf{K}$ and $\bf{Q}$ minima. Even if this observation is qualitatively in line with previous reports for homogeneously doped monolayer MoS$_2$\cite{PhysRevB.87.241408}, the phonon softenings are much less pronounced \cite{PhysRevB.87.241408,PhysRevB.90.245105}, once again confirming the importance of employing the proper FET setup in the calculation and a number of layers large enough to be representative of the thicker samples used in experiment (see also the Appendix). 
The inclusion of anharmonic effects further reduces the phonon softening predicted within the harmonic approximation. The effect of anharmonicity is increasingly important as the doping increases, and becomes extremely important in the high doping regime ($\geq$ 1.4 $\times 10^{14}$ e$^-$/cm$^{2}$, red and green curves in Fig. \ref{fig11} a). The sizable effect of anharmonicity at high doping can be well appreciated in Fig. \ref{fig13}, where we compare the harmonic (yellow) and anharmonic (blue) phonon dispersion at $n_e=$2.01$\times$ 10$^{14}~e^-$/cm$^{2}$. We find that at this doping the correction can be as large as 10-30\% of the harmonic phonon frequency value across the softer acoustic phonon branch.  

\begin{figure}[htpb!]
\includegraphics[width=1\linewidth]{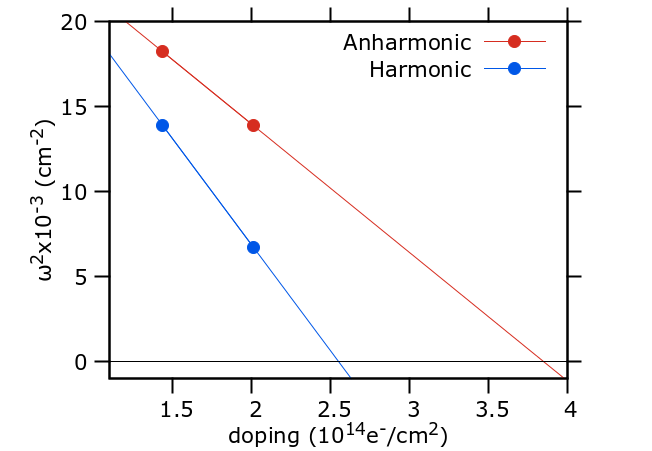}
    \caption{Predicted CDW instability at the M-point of the BZ as a function of electron doping with (red) and without (blue) the anharmonic correction.}\label{fig12}
\end{figure}

\begin{figure}[htpb!]
\includegraphics[width=1\linewidth]{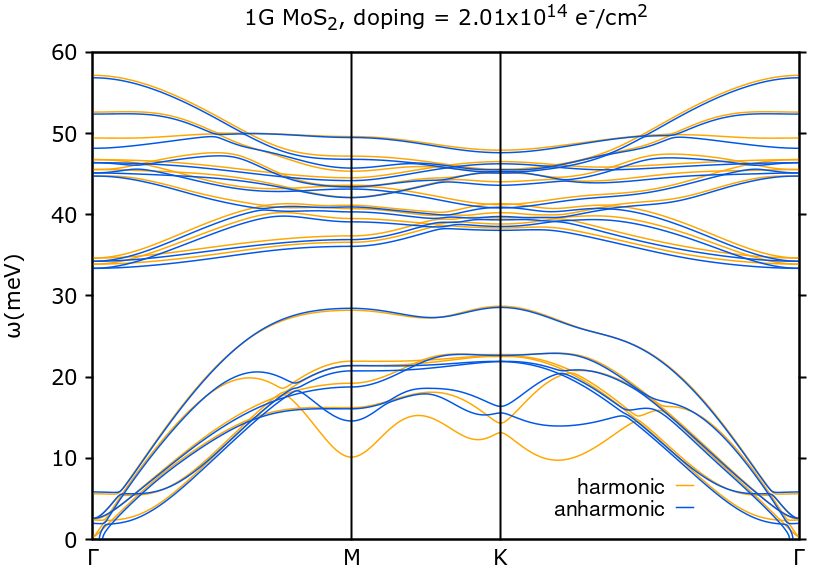}
    \caption{Phonon dispersion for single-side gated MoS$_2$ at $n = 2.01 \times$ 10$^{14}~e^-$/cm$^{2}$ calculated within the harmonic approximation (yellow) and including the anharmonic contribution to the phonon dispersion  (blue). }\label{fig13}
\end{figure}

\begin{figure*}[t]
\centering
\includegraphics[width=0.8\linewidth]{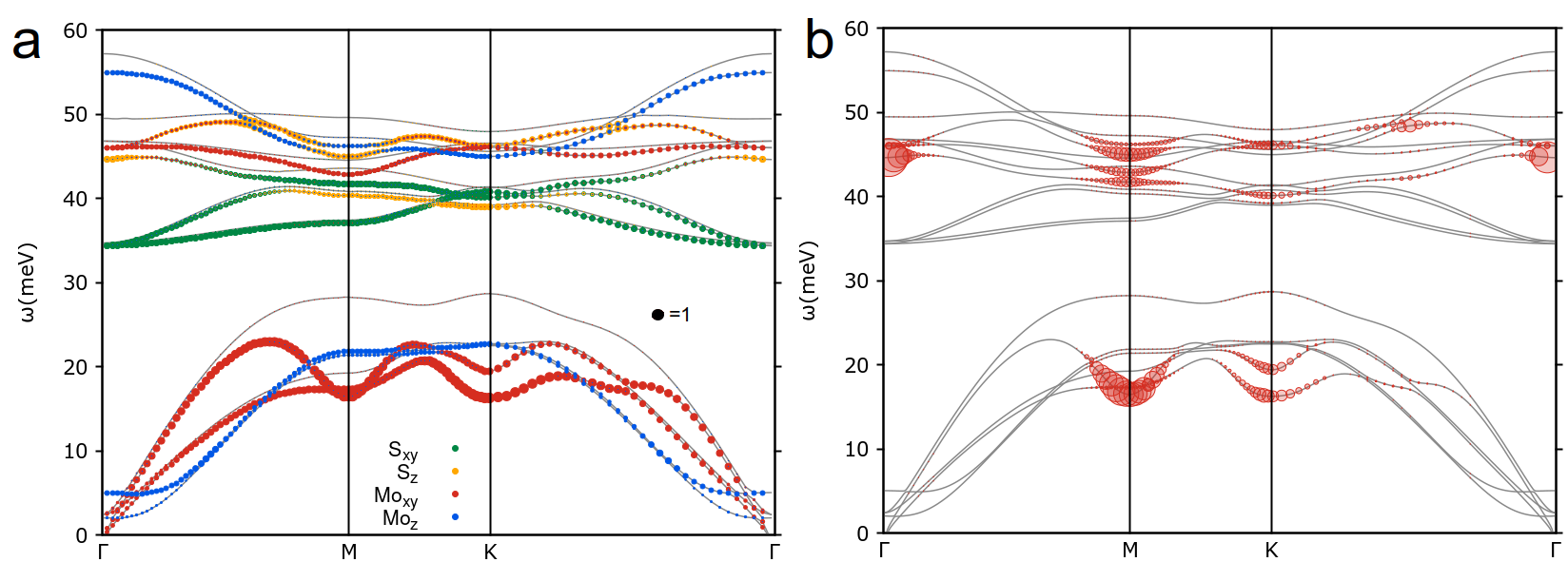}
    \caption{Panel a: phonon frequencies for single-side gated MoS$_2$ at $n_e=$1.15 $\times$ 10$^{14}~e^-$/cm$^{2}$. The size of the circles is proportional to the projection of the phonon eigenvector onto the Mo$_{xy}$,Mo$_{z}$,S$_{xy}$,S$_{z}$ for the outer layer (closer to the gate),respectively. The size of the black circle size indicates a unitary projection onto a single component $\epsilon_{\mu}^{s \alpha}$ for scale. Panel b: phonon dispersion for single-side gated MoS$_2$ at $n_e=$1.15 $\times$ 10$^{14}~e^-$/cm$^{2}$. The diameter of red circles is equal to 1/4 $\gamma_{{\bf q}\nu}$ from Eq.\ref{eq:gamma_Allen} (i.e. the full-width half-maximum divided by 4).}\label{fig14}
\end{figure*}

In previous works\cite{PhysRevB.90.245105} based on uniform doping of single layer MoS$_2$, it was claimed that a CDW occurs in the doping region relevant for superconductivity. However, this is an artifact of simulating a single layer in the uniform doping setup and neglecting anharmonicity. As shown in Fig. \ref{fig12}, for thicker layers, the simulations with explicit inclusion of anharmonicity and FET setup pushes the occurrence of CDW to very high doping, namely close to $n_e\approx 4\times 10^{14} {\rm e}^- /{\rm cm}^2$, a doping that is too large to be relevant for experiments on single gate MoS$_2$. Thus, the occurrence of a charge density wave instability cannot explain the flattening of the superconducting critical temperature seen in experiments.

In Fig. \ref{fig14} a we project the phonon eigenvectors  onto their atomic components, namely $\epsilon_{\mu}^{s \alpha}(\mathbf{q})$, where $s$ is the atom index and $\alpha$ the cartesian component, while $\mu$ labels the phonon mode. We observe that the main atomic character for the low-energy  modes at the $M$ point arises from molybdenum atoms. We also plot in Fig.  \ref{fig14} b the phonon linewidth (full width half maximum), namely for the mode $\nu$ with phonon momentum ${\bf q}$ the quantity:

\begin{equation}
\label{eq:gamma_Allen}
\gamma_{{\bf q}\nu} = \frac{4\pi \omega_{{{\bf q} \nu}}}{N_k} \sum_{{\bf k},n,m} |g_{{\bf k}n,{\bf k}+{\bf q}m}^{\nu}|^2 \delta(\varepsilon_{{\bf k}n}) \delta(\varepsilon_{{\bf k+q}m})
\end{equation}
where  $g_{{\bf k}n,{\bf k}+{\bf q}m}^{\nu}=\sum_{s\alpha}
\epsilon^{s\alpha}_{\nu}({\bf q}) d_{{\bf k}n,{\bf k}+{\bf q}m}^{s\alpha}/
{\sqrt{2M_s\omega_{{\bf q}\nu}}}$ is the electron-phonon matrix element.
The matrix element  $d^{s\alpha}_{{\bf k}n,{\bf k}+{\bf q}m} = \langle{{\bf k}n}|
\delta V_{KS} / \delta u^{s\alpha}(\mathbf{q}) |{{\bf k}+{\bf q}m}\rangle$
is the deformation potential, $V_{KS}$ is the Kohn-Sham potential, $|{\bf k}n\rangle$ is a
Kohn-Sham state with energy $\epsilon_{{\bf k}n}$
measured from the Fermi  level ($\epsilon_F$), and $N_k$ is the number of k-points used in the calculation. Phonon frequencies are labeled $\omega_{\mathbf{q}\nu}$. Finally $M_s$ is the mass of the $s^{\rm th}$ ion. It is worthwhile to recall that the phonon linewidth does not depend on the phonon frequency.

The relation between the electron-phonon coupling $\lambda_{{\bf q},\nu}$ of a given mode $\nu$ at a phonon momentum ${\bf q}$ and linewidth was first pointed out by Allen\cite{PhysRevB.6.2577}, namely:

\begin{eqnarray}
\lambda_{{\bf q},\nu}=\frac{\gamma_{{\bf q}\nu}}{2\pi N(0)\omega_{{\bf q}\nu}^2}
\label{lambda}
\end{eqnarray}
where $N(0)$ is the density of states per spin at $\epsilon_F$. 

From Fig. \ref{fig14} a, it is clear that the main contribution to the phonon linewidth (and consequently to the electron-phonon interaction) arises from the lowest energy phonon band, in particular from the phonon softening at the M-point. Indeed, the $\bf{M}$ wavevector corresponds to the nesting between the $\bf{K}$ and $\bf{Q}$ valleys.  Other notable contributions arise from the the $\bf{K}$ wavevector (nesting different $\bf{K}$ valleys) and from one optical mode at $\Gamma$. This result shows that the electron-phonon interaction in single-side gated MoS$_2$ has a dominant inter-valley character. 
Furthermore,
by comparing Figs.\ref{fig14} a and b, we conclude that the most coupled modes have a dominant in-plane molybdenum character.

In order to gain insight on the electron-phonon interaction we calculate the Eliashberg function, namely

\begin{eqnarray}
&& \alpha^{2}F(\omega) = \frac{1}{N(0) N_k N_q} \sum_{{\bf k}{\bf q}nm}
  \sum_{st \alpha \beta \mu}
  \frac{\epsilon_{\mu}^{s \alpha}(\mathbf{q})
           \epsilon_{\mu}^{t \beta *}(\mathbf{q}) }{
        2 \omega_{\mathbf{q}\mu}  \sqrt{M_sM_t}} \nonumber \\
&& \ \times
   d^{s\alpha}_{{\bf k}n,{\bf k}+{\bf q}m}
   d^{t\beta *}_{{\bf k}n,{\bf k}+{\bf q}m}
   \delta(\epsilon_{{\bf k}n})
   \delta(\epsilon_{{\bf k+q}m})
   \delta(\omega -  \omega_{\mathbf{q}\mu}),
\label{eliashberg}
\end{eqnarray}

where $N_q$ are the number of 
 phonon momentum points used for the BZ sampling.
The Eliashberg function can be calculated in the
harmonic or anharmonic case if the harmonic or SSCHA
phonon frequencies and polarizations are used
in Eq. \eqref{eliashberg}. 

In Fig. \ref{fig11} b we plot the Eliashberg function (filled curves)
as a function of doping together with the cumulative integral $\lambda(\omega) = 2~\int_{0}^{\omega} dy~\alpha^2F(y)/y$ (dashed lines). Despite the fact that the electron-phonon coupling strength does not vary appreciably in the considered doping range (the first momentum of the Eliashberg function is essentially constant), the progressive softening of phonon frequencies enhances $\lambda(\omega)$ as a function of doping. 

\begin{figure}[h]
\includegraphics[width=0.9\linewidth]{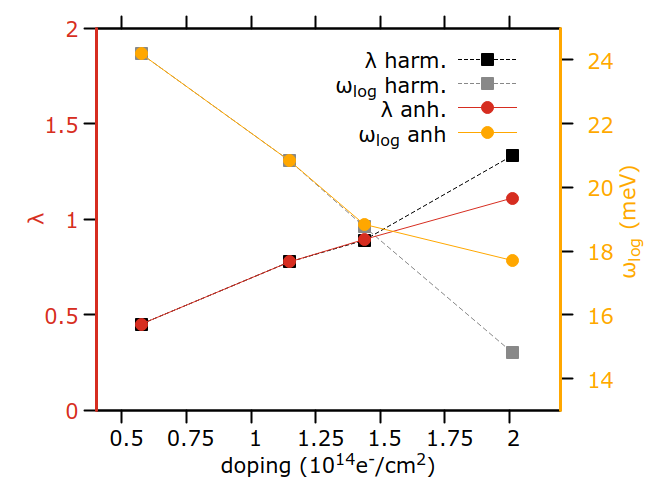}
    \caption{Electron phonon coupling parameter $\lambda$ (red) and logarithmic average phonon frequency (yellow) as a function of electron doping for single-side gated MoS$_2$, with the inclusion of anharmonic correction. Black and grey squares represent the same quantities without including anharmonic corrections.}\label{fig15}
\end{figure}

In Fig. \ref{fig15}, we plot the electron-phonon coupling parameter $\lambda=\lambda(\infty)=\frac{1}{N_q}\sum_{{\bf q},\nu}\lambda_{{\bf q}\nu} $ as a function of doping, both in the harmonic approximation (black squares) and with the inclusion of the anharmonic contributions in the phonon frequencies and eigenvectors (red circles).  We observe a progressive increase of $\lambda$ as a function of doping, partially mitigated by anharmonic effects. The yellow curve in the same figure displays the behavior of the logarithmic average phonon frequency $\omega_{log}$ as a function of doping, which is progressively reduced due to phonon softening. 

As the calculated 8.3\% increase of $N(0)$ in single-gated MoS$_2$ from 0.57 to 1.72 $\times 10^{14}$ e$^-$/cm$^{2}$ has a weak effect on $\lambda$, we can safely conclude that phonon softening is the main contribution to electron-phonon coupling enhancement, as $\lambda$ is enhanced by a factor of 2 in the same range.   

\begin{figure}[h]
\includegraphics[width=0.8\linewidth]{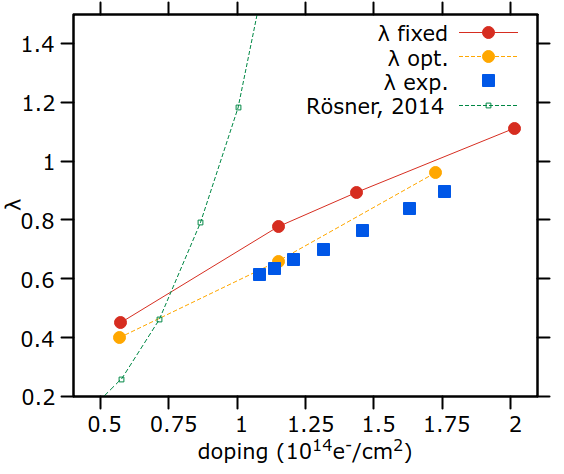}
    \caption{Calculated electron phonon coupling parameter $\lambda$ with fixed (red) and optimized (yellow) in-plane lattice parameter, compared to experimental data of Ref.\cite{AliElYumin2019} (blue squares). Green points represent previous calculations from Ref.\cite{PhysRevB.90.245105}.}\label{fig16}
\end{figure}

In Fig. \ref{fig16} we compare the calculated electron-phonon coupling parameter $\lambda$ to the one extracted from resistivity measurements in the normal state in  Ref.\cite{AliElYumin2019}. Here, we both consider the case where the in-plane lattice parameter is kept fixed (red circles) and the case where the in-plane lattice parameter is optimized at each doping, as explained in Sec.\ref{sec2}. There is a a remarkable quantitative agreement between theory and experiments, especially for the case where the lattice parameter is optimized. Most important, the dependence of the electron-phonon coupling on doping is perfectly reproduced. 

We underline that at a doping of $n_e=1.38\times 10^{14}$ e$^-/{\rm cm}^{2}$, the electron-phonon coupling for a single layer MoS$_2$ with uniform doping was evaluated to be $\lambda=8$ in Ref. \cite{Roesner2017}. This is an overestimation of a factor of $11.4$ with respect to our result and with respect to experiments. This dramatic overestimation underline the need of properly including all the relevant effects (field-effect setup, thickness, anharmonicity) to correctly evaluate the electron-phonon interaction in FET doped dichalcogenides.

 Finally, in order to quantify the role of inter-band and the intra-band contributions in determining the average electron-phonon coupling $\lambda$, we decompose the electron-phonon calculation parameters in two contributions, namely we write $\lambda = \lambda_{intra} + \lambda_{inter}$, where

\begin{equation}
\begin{gathered}
        \lambda_{intra} = \sum_{\mathbf{q}\leq\mathbf{q_{coff}},\nu} \lambda_{\mathbf{q},\nu}\\
        \lambda_{inter} = \sum_{\mathbf{q}>\mathbf{q_{coff}},\nu} \lambda_{\mathbf{q},\nu}\\
\end{gathered}
\end{equation}

We find $\lambda_{intra}$ = 0.015 and $\lambda_{inter}$ = 0.755 at $n_e$ $=$ 1.15$\times$ 10$^{14}~e^-$/cm$^{2}$, indicating that the electron-phonon coupling is almost totally due to inter-band electron-phonon scattering.

\begin{figure}[h]
\includegraphics[width=1\linewidth]{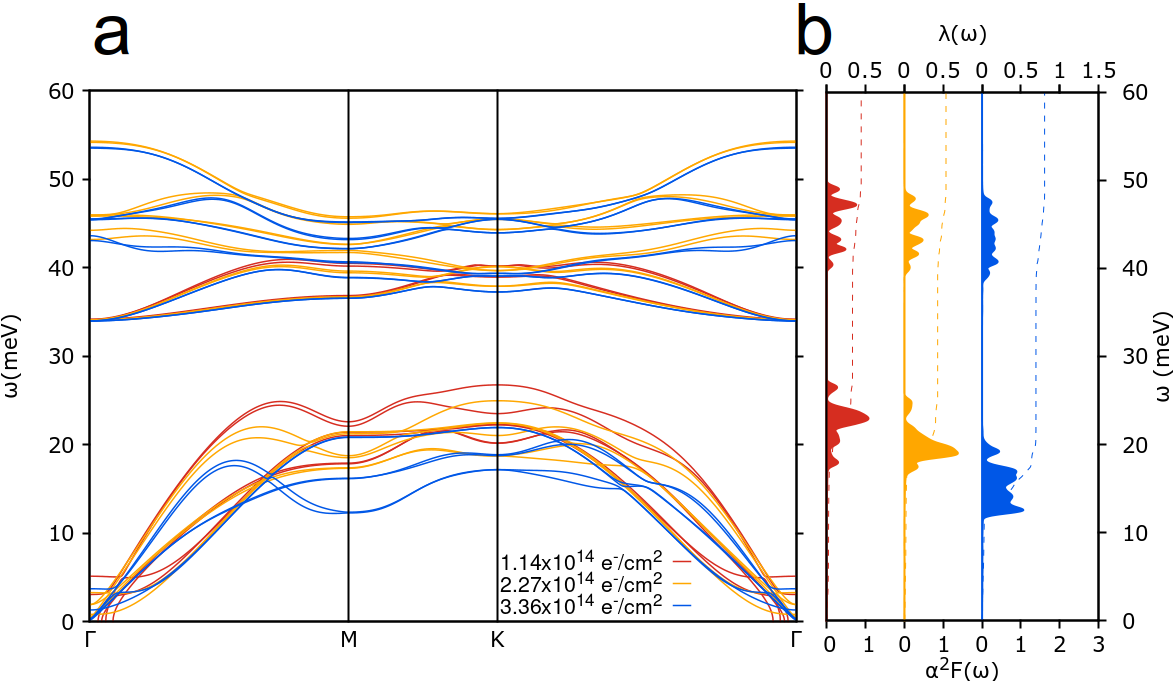}
    \caption{Panel a: phonon dispersion as a function of electron doping for double-side gated MoS$_2$. Panel b: corresponding Eliashberg function $\alpha^2F(\omega)$ (filled curves) and $\lambda({\omega})$ (dashed lines).}\label{fig17}
\end{figure}

\begin{figure}[h]
\centering
\includegraphics[width=0.9\linewidth]{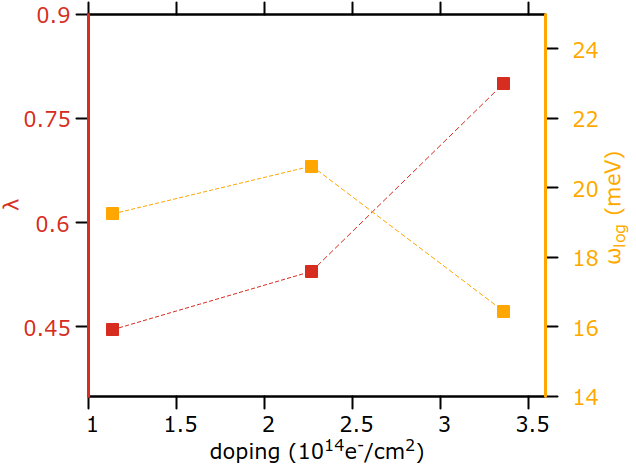}
    \caption{Electron phonon coupling parameter $\lambda$ (red) and logarithmic average phonon frequency (yellow) as a function of electron doping for double-side gated MoS$_2$.}\label{fig18}
\end{figure}

\begin{figure}[htpb!]
\centering
\includegraphics[width=1\linewidth]{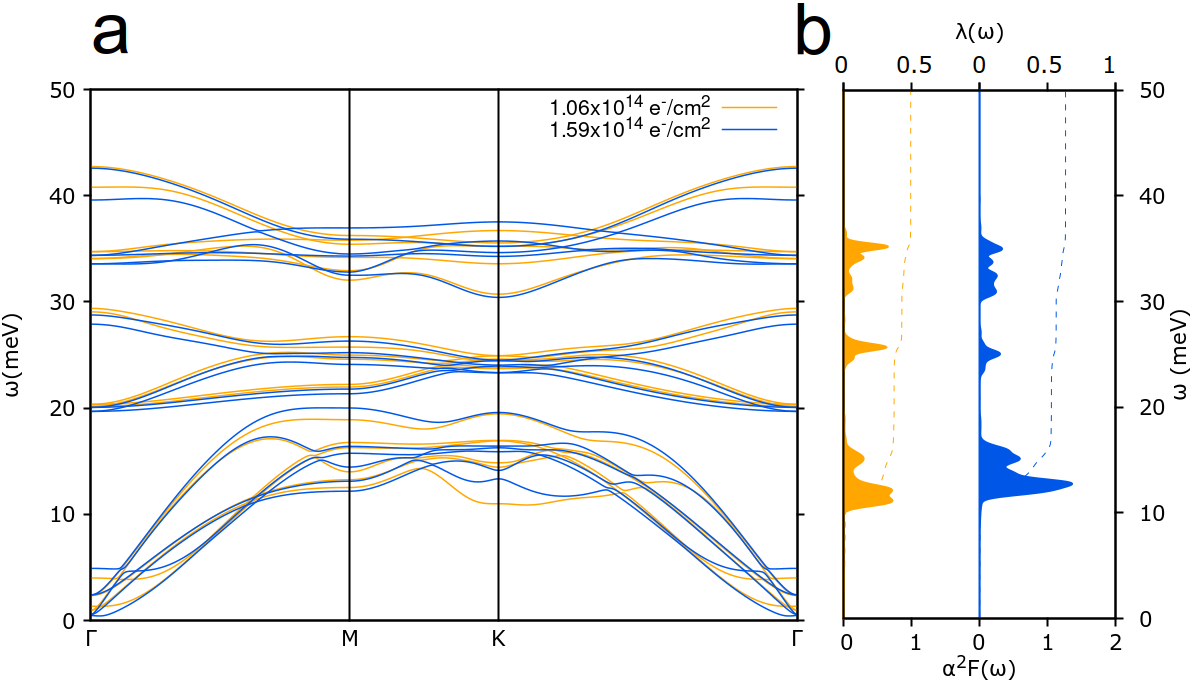}
    \caption{Panel a: phonon dispersion as a function of electron doping for single-side gated MoSe$_2$. Panel b: corresponding Eliashberg function $\alpha^2F(\omega)$ (filled curves) and $\lambda({\omega})$ (dashed lines). Anharmonic effects are included at 1.59$\times$ 10$^{14}~e^-$/cm$^{2}$.}\label{fig19}
\end{figure}

\subsection{Double-side gated MoS$_2$}

We now consider the case of double-side gated MoS$_2$. 
In Fig. \ref{fig17} we plot the harmonic phonon dispersion (panel a) and the Eliashberg function. We consider the case with the optimized in-plane lattice parameter as a function of doping. The effects of keeping a fixed lattice parameter on the superconducting properties (or equivalently of strain) are discussed in the next section. 

The lowest energy phonon bands are generally higher in energy than those of a single gate MoS$_2$ and remain well above $10$ meV in the considered doping range, and thus we expect anharmonic correction to be relatively small. For this reason we did not perform the anharmonic calculation in this system.

The higher phonon frequencies are related to a generally lower electron-phonon coupling as a function of doping with respect to the single-gated case, as observed in Fig. \ref{fig18}, where we plot $\lambda$ and $\omega_{log}$  for double-side gated MoS$_2$. The electron-phonon coupling parameter $\lambda$ becomes sizable at very high doping values of the order of $n_e$ = 3~$\times$~10$^{14}~e^-$/cm$^{2}$, while the logarithmic average phonon frequencies becomes sensibly lower ($\approx$ 16 meV). These results are compatible with the absence of superconductivity at low doping in experiments and with the lower T$_c$ of double gated MoS$_2$ with respect to the single gated case. Differently from the single-side gated case, here $N(0)$ sensibly increases in the studied doping range (30\% increase from  1.14 to 3.36 $\times 10^{14}$ e$^-$/cm$^{2}$) and plays an important role in the increase of $\lambda$. This effect cooperates with  the decrease of the logarithmic phonon frequency $\omega_{log}$ as a function of doping in increasing T$_c$. 

\begin{figure}[htpb!]
\centering
\includegraphics[width=0.8\linewidth]{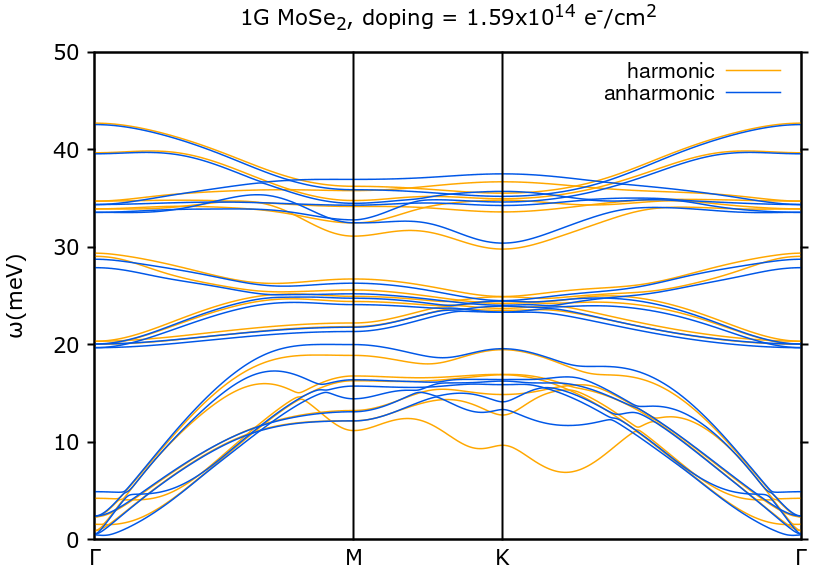}
    \caption{Phonon dispersion for single-side gated MoSe$_2$ at $n = 1.59\times$ 10$^{14}~e^-$/cm$^{2}$ calculated within the harmonic approximation (yellow) and including the anharmonic contribution to phonon frequencies within SSCHA (blue).}\label{fig20}
\end{figure}

\begin{figure}[htpb!]
\centering
\includegraphics[width=0.9\linewidth]{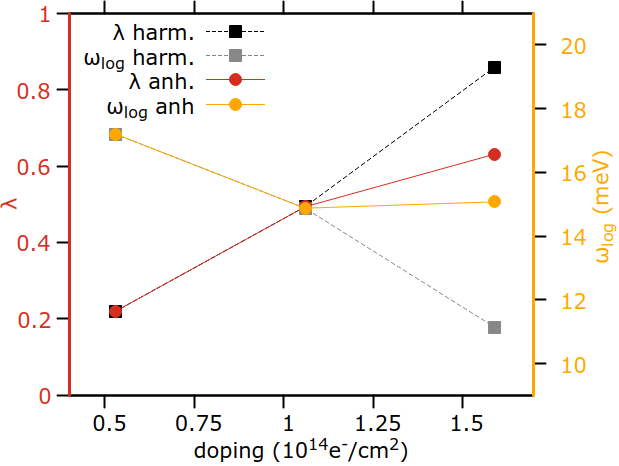}
    \caption{Electron phonon coupling parameter $\lambda$ (red) and logarithmic average phonon frequency (yellow) as a function of electron doping for single-side gated MoSe$_2$. Black and grey squares represent the same quantities calculated without including anharmonic corrections.}\label{fig21}
\end{figure}

\begin{figure*}[t!]
\centering
\includegraphics[width=0.8\linewidth]{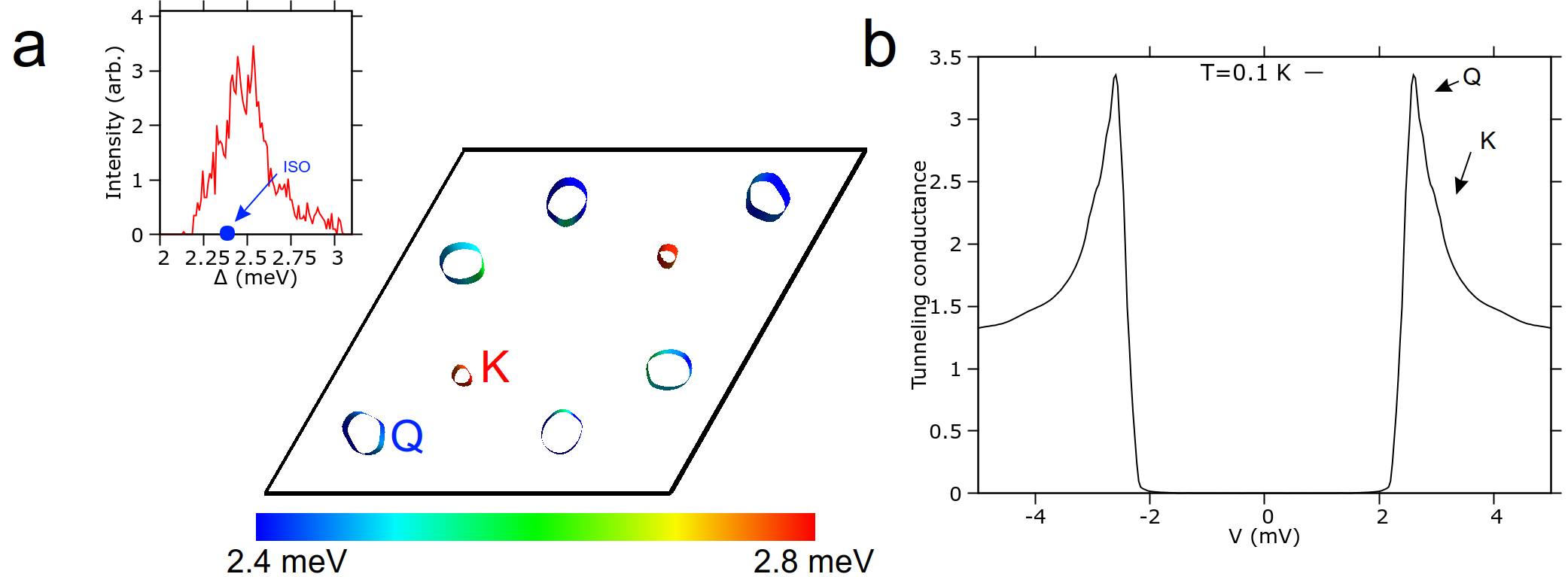}
    \caption{Panel a: 2D Fermi surface for single-side gated MoS$_2$ at 1.44 $\times 10^{14}$ e$^-$/cm$^{2}$ (spin-orbit coupling is not included). Color is proportional to the superconducting gap calculated solving the Migdal-Eliashberg equation at 1.5~K. Plot realized using FermiSurfer\cite{KAWAMURA2019197}. Inset: anisotropic density of states for the Migdal-Eliashberg gap as a function of the energy. Panel b: calculated tunneling spectrum for single-side gated MoS$_2$ at 1.44 $\times$10$^{14}~e^-$/cm$^{2}$ and T=0.1~K.}\label{fig22}
\end{figure*}

\subsection{Single-side gated MoSe$_2$}

Finally we discuss the case of single-side gated MoSe$_2$. In this system in the doping range relevant for experiments only the minimum at $\mathbf{Q}$ is populated. As the large electron-phonon interaction and the consequent softening at the point ${\bf M}$ are related to inter-valley ${\bf K}$-${\bf Q}$ scattering, we expect the softening at ${\bf M}$ not to occur. 

Indeed, the phonon dispersion as a function of doping is completely different, as the  ${\bf K}$-${\bf Q}$ scattering channel is not present (Fig. \ref{fig19}). Nevertheless, we find that for the highest considered doping ($n_e$~=~$1.59~\times$~10$^{14}~e^-$/cm$^{2}$) anharmonicity strongly affects phonon frequencies, as depicted in Fig. \ref{fig20}, where we compare the harmonic and anharmonic phonon spectrum. We find that increasing the doping concentration from $1.06~\times$~10$^{14}~e^-$/cm$^{2}$ to $1.59~\times$~10$^{14}~e^-$/cm$^{2}$ induces  a prominent softening in the $\bf{K}$-$\bf{\Gamma}$ direction at the harmonic level, related to the $\bf{Q}$-$\bf{Q}$$'$ inter-band scattering (nesting at approx. $\frac{1}{2}\bf{\Gamma}\bf{K}$). A similar softening can also be observed for single-side gated MoS$_2$ in Fig. \ref{fig11} a at high doping. 

The inclusion of anharmonicity largely reduce the softening, operating a correction as large as 40\% of the harmonic frequency to the softer acoustic branch (see Fig. \ref{fig20}). 

While we expect a weaker superconductivity in single-side gated MoSe$_2$ with respect to single-side MoS$_2$ case due to the absence of $\bf{K}$-$\bf{Q}$ inter-band scattering, the high $\lambda$ value of 0.63 calculated at 1.59~$\times$~10$^{14}~e^-$/cm$^{2}$ (Fig. \ref{fig21})  still corresponds to a sizable superconducting critical temperature T$_c$. Similarly to the case of single-gated MoS$_2$, the $N(0)$ increase is marginal (9\% increase from 0.53 to $1.59 \times$ 10$^{14}~e^-$/cm$^{2}$).

\section{Superconductivity}\label{sec6}

We calculate the superconducting properties using three different approaches, namely anisotropic multiband Eliashberg equations, isotropic Eliashberg equations and the Allen and Dynes/McMillan formula. In all these approaches a crucial ingredient is the treatment of the electron-electron repulsion. We employ the Morel-Anderson pseudopotential\cite{PhysRev.125.1263} to parametrize the Coulomb repulsion in the superconducting state. For a single layer MoS$_2$, a value of $\mu^*=0.13$ has been initially proposed\cite{PhysRevB.90.245105}. Subsequent studies  demonstrated that while the diagonal intra-valley repulsion is correctly described by this value\cite{PhysRevB.94.134504}, the inter-valley Coulomb repulsion is sensibly lower. Since we find that most of the electron-phonon coupling in this system stems from inter-valley coupling ($\bf{K}$-$\bf{Q}$, $\bf{K}$-$\bf{K}$$'$ and $\bf{Q}$-$\bf{Q}$$'$), we decided to use a lower value for the Coulomb pseudopotential equal to $\mu^*=0.1$, in an attempt to mimic the reduced repulsion affecting the inter-valley Cooper pairs. 
In principle the value of $\mu^{*}$ depends on the system. Here, we keep the same value for all systems to avoid introducing additional fitting parameters.

In Fig. \ref{fig22} a, we plot the Fermi surface of single-side gated MoS$_2$ at 1.44 $\times$10$^{14}~e^-$/cm$^{2}$, colored according to the value of the superconducting gap. A substantial anisotropy is present. It is interesting to note that the average superconducting gap at the $\bf{K}$-minimum is larger than the one at the $\bf{Q}$-minimum, thus single-side gated MoS$_2$ can be effectively considered a multi-gap superconductor. However, no clear double gap is identified in the superconducting gap spectrum (see inset of Fig. \ref{fig22}, red curve), due to the small energy separation existing between the two gaps. 

In Fig. \ref{fig22} b, we show the corresponding tunneling spectrum at T = 0.1 K. As it can be seen, the presence of two shoulders in the tunneling spectrum is  associated to gap anisotropy.

\begin{figure*}[htpb!]
\centering
\includegraphics[width=0.8\linewidth]{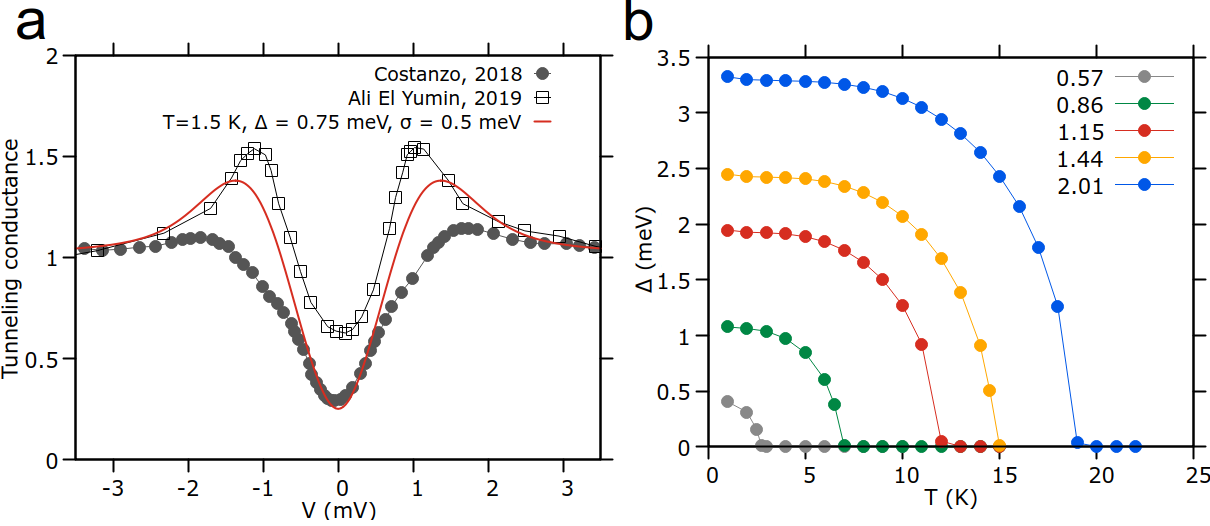}
    \caption{Panel a: calculated tunneling spectrum for single-side gated MoS$_2$ considering a gap $\Delta = 0.75$ meV and T=1.5~K, including a Gaussian disorder with $\sigma = 0.5$~meV (red line), compared with measured tunneling spectrum from Ref.\cite{Costanzo2018} at T=1.5~K and doping 1.5~$\times$10$^{14}~e^-$/cm$^{2}$ (filled circles) and from Ref.\cite{AliElYumin2019} at T=1.45~K and doping 1.8$\times$10$^{14}~e^-$/cm$^{2}$. Lines connecting experimental points are a guide to the eye. Panel b: solution of the isotropic Eliashberg equation as a function of doping for MoS$_2$. Here, a constant $\mu^* = 0.1$ is assumed and doping is expressed in 10~$\times^{14}e^-/$cm$^2$.}\label{fig23}
\end{figure*}

We now discuss our results in the light of experimental tunneling measurements\cite{Costanzo2018,AliElYumin2019}. 
Before proceeding it is however important to stress that for a given T$_c$ the experimental measurements report substantially different superconducting gaps, both in magnitude and in shape. For examples, in Ref. \cite{Costanzo2018} for a T$_c=10$ K the measured energy 
gap ranges from 1 to 2 meV depending on the sample (a variation of a factor of $2$).

This is even more clearly exemplified in  Fig. \ref{fig23} a, where we report the measured gap in Ref.\cite{Costanzo2018} at $n_e$~=~$1.5~\times$~10$^{14}~e^-$/cm$^{2}$ (T = 1.5 K) (filled circles) and at $n_e$~=~$1.8~\times$~10$^{14}~e^-$/cm$^{2}$ (T = 1.45 K) in Ref.\cite{AliElYumin2019} (empty squares). It appears clear that the two experimental reports show substantial differences despite the formally identical doping, both in the observation of one or two features in the tunneling spectrum and in the amount of tunneling conductance suppression due to superconductivity. However, it appears evident that the experimentally measured gap for these two samples is smaller than the one that we calculate at 1.44 $\times~10^{14}$ e$^-$/cm$^{2}$ (Fig. \ref{fig22} b). Due to the large variability of the gaps measured in experiments for the same T$_c$ we rescale our gap to the experimental one  (average gap, $\Delta = 0.75$ meV), and introduce other effects  causing a smearing the tunneling spectrum. Some possibilities are represented by a finite linewidth\cite{AliElYumin2019}, or charge inhomogeneity\cite{Costanzo2018}. This effects can be  describes by a Gaussian distribution with standard deviation $\sigma = 0.5$~meV. (Fig. \ref{fig23} a, red curve). Given the large variability of the experimental tunneling spectra (larger than the difference between theory and experiment) it is difficult to assess the accuracy of the theoretical approaches in determining the superconducing gap.

In Fig. \ref{fig23} b, we calculate the superconducting gap for single-gated MoS$_2$ as a function of doping and temperature. Here, the isotropic  Eliashberg equations are employed, as we verified that neglecting anisotropic effects does not substantially alter neither the critical temperature nor the average gap. For example, at 1.44 $\times$ 10$^{14}~e^-$/cm$^{2}$, the isotropic prediction is $\approx$ 2.4 meV while the average anisotropic gap is $\approx$ 2.5 meV (see also inset in Fig. \ref{fig22}). 
Globally the $2\Delta/k_B T_c$ ratio is very close to
the BCS value.

\begin{figure}[htpb!]
\includegraphics[width=1\linewidth]{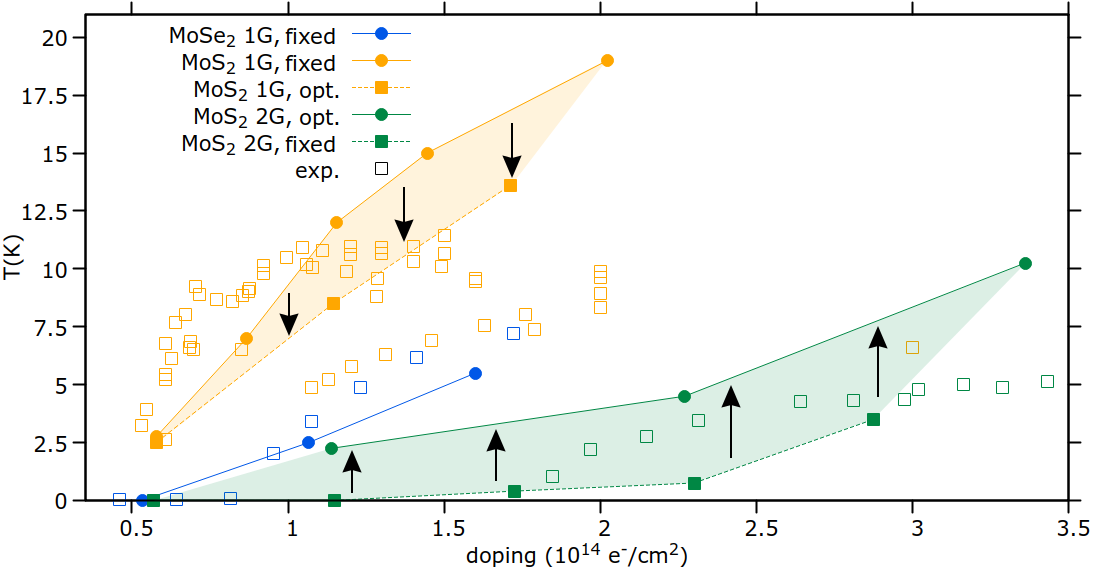}
    \caption{Superconducting critical temperature for MoSe$_2$ (blue) and MoS$_2$ (yellow) as a function of doping, employing a constant $\mu^*  = 0.1$. Filled points represent theoretical values calculated solving the anisotropic Eliashberg equations; empty squares represent experimental data from Refs.\cite{Costanzo2018,doi:10.1126/science.1228006,PMID:26235962,Saito2016,doi:10.1126/science.aab2277,AliElYumin2019,Zheliuk2019}. The arrows mark the effect of in-plane structural optimization on the critical temperature.}\label{fig24}
\end{figure}

We extend the analysis of superconductivity to the case of single-side gated bilayer MoSe$_2$ and double-side gated bilayer MoS$_2$. In Fig. \ref{fig24} we summarize the calculated T$_c$ with $\mu^*=0.1$ for all systems and compare  to experimental data. The shaded region represents the uncertainty that we associate to the use of an optimized  or fixed  lattice parameter. Indeed, it is not clear if the best way to simulate experiments is to constrain or not the in-plane lattice parameter as this depends on the substrate on which the sample is deposed and if it is allowed to expand. The arrows mark the effect of the in-plane structural optimization.

For the single-gated  MoS$_2$ case, our predictions display a globally good agreement with experiments, although notably no dome trend is observed in our calculations, rather an ever increasing T$_c$ is reported. Furthermore, we predict that optimizing the lattice parameter as a function of doping causes an expansion (0.85\% in-plane parameter expansion at $n_e=$1.15 $\times$ 10$^{14}~e^-$/cm$^{2}$), corresponding to a critical temperature reduction from 12 to $\approx$ 7 K (at at $n_e=$1.15 $\times$ 10$^{14}~e^-$/cm$^{2}$). The sizable change in the superconducting T$_c$ can be traced back to the low energy band structure of single-side gated MoS$_2$, in particular to the fact that $\bf{K}$ and $\bf{Q}$ minima position are strongly dependent on the structural parameters.

We note that the predicted T$_c$ for the MoSe$_2$ case is in very good agreement with expeeriments and leads a lower T$_c$ than in the MoS$_2$ case, as expected from experiments.
We verified that the inclusion of in-plane optimisation in MoSe$_2$ does not qualitatively alter the minima occupation (i.e. only the $\bf{Q}$ minimum is occupied also considering in-plane expansion), and, thus, no appreciable effects on T$_c$ are seen.
However, it is important to remark that even in the case of single gate MoSe$_2$, theoretical calculations predict no T$_c$ saturation with doping but an increasing T$_c$. 

Finally, for the case of a double-gated bilayer MoS$_2$, we estimate a lower critical temperature with respect to the single-gated MoS$_2$ and MoSe$_2$ case, again in agreement with experiments\cite{Zheliuk2019}. 
We note a general overestimation of T$_c$ for double-side gated MoS$_2$ when using the optimized parameter (green filled circles). In particular, the predicted early onset of superconductivity in our calculations for double-gated MoS$_2$ with respect to the experiments could be explained by the delicate dependence of the electronic structure on the considered structural parameters. Indeed,  we observe that superconductivity is essentially absent in the low doping regime in the absence of strain, even if the strain is as small as 0.5\% at 1.14 $\times$ 10$^{14}~e^-$/cm$^{2}$. This is because only the $\bf{Q}$ minimum is occupied in the absence of tensile strain. This delicate dependence of the minima occupation from the considered structural parameter suggests that even a very small compressive strain in double-side gated MoS$_2$ could explain the anticipated onset of superconductivity in our calculations with respect to the experiments. 

The capability of predicting this trend among the different dichalcogenides and the different geometries demonstrate the reliability of our theoretical approach.

\begin{table}
    \centering
    \begin{tabular}{|c|c|c|c|}
    \hline
  $n_e$($\times 10^{14}$ e$^-$/cm$^{2}$) & $\mu^*$ & T$_c$ \\
 \hline
 \hline
     0.57 & 0.1 & 2.75 \\
    1.15 & 0.1 &  7 \\
    1.44 &
    0.1  &  12\\ 
    1.72 & 0.15   & 11.5 \\ 
    \hline
    \end{tabular}
    \caption{Values of the Coulomb pseudopotential parameter for single-side gated MoS$_2$ to reproduce T$_c$ trend as a function of doping in Fig. \ref{fig24} (red circles).}
    \label{tab2}
\end{table}

\begin{figure}[h]
\includegraphics[width=1\linewidth]{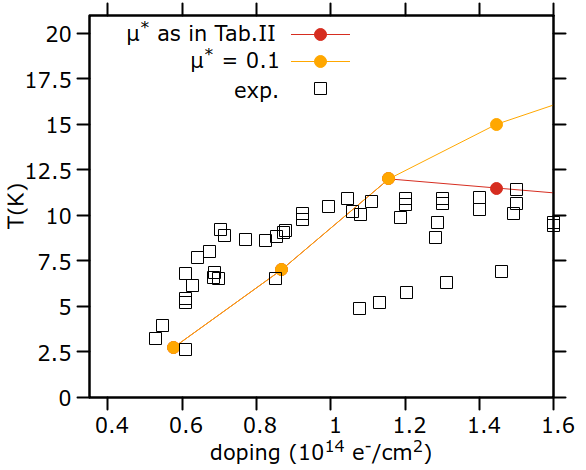}
    \caption{Superconducting critical temperature for MoSe$_2$ (blue) and MoS$_2$ (yellow) as a function of doping, employing a constant $\mu^*  = 0.1$. Full circles represent theoretical values calculated solving the anisotropic Eliashberg equation; empty squares represent experimental data from Refs.\cite{Costanzo2018,doi:10.1126/science.1228006,PMID:26235962,Saito2016,doi:10.1126/science.aab2277}. Red circles represent the superconducting critical temperature for single-side gated MoS$_2$ employing values of $\mu^*$ as in Tab.\ref{tab2}.}\label{fig25}
\end{figure}

However, we have to underline that, while there is a global remarkable agreement between theoretical  and experimental
superconducting critical temperatures in all
compounds and in all FET geometries, a notable difference resides in the absence of saturation in the T$_c$ versus doping curves. Indeed, in all case, if calculations are carried out at doping larger than the those achieved in experiments, it is found that T$_c$ increases without saturation (or with a very weak saturation). The lack of saturation is intrinsic in all these compounds and depends on several facts namely, the increase in the density of states due to band-structure  deviations from parabolicity (see Figs. \ref{fig6} and \ref{fig9} )  or, as in the case of double side gated MoS$_2$,
the occupation of additional minima at $\bf{Q}$ (see Fig.  \ref{fig8}), generating a sudden increase in T$_c$.
Our calculations indeed suggest that in the absence of other extrinsic effects the superconducting T$_c$ could be made even larger in FET doped molybdenum dichalcogenides.

\section{Discussion on T$_c$ saturation\label{sec:disorder}}

The question naturally arises as to why there is a saturation of T$_c$ versus doping in these systems.

Some insights can be obtained analyzing a basic experimental fact, namely  the extreme variability of the measured superconducting gap for different samples having the same T$_c$ (in some case 100\% variation). This suggests that probably the superconducting gap is extremely inhomogeneous in the sample and if it were possible to measure the superconducting gap in different points of the sample, as in an STS experiment, very different values would be found at different point on the surface.
A possible reason for this could be the very non-uniform nature of the doping in the dichalcogenides layer (i.e. in the plane, not across the sample). This is indeed highly likely due to the  freezing of the electrolyte at low temperature inducing charge pockets in different points in the sample and consequently different regions with different doping.

This conjecture is corroborated by recent measurements of T$_c$ in single-gated single-layer WS$_2$ \cite{doi:10.1073/pnas.1716781115} showing the occurrence of a T$_c$ dome feature ultimately resulting in a reentrant insulating state at high doping that was interpreted as originating from the occurrence of highly disordered charge carrier inhomogeneities poorly screened in the 2D limit. 

 The localization of doped charges (carriers) could then be the  precursor to an Anderson localization, resulting in an enhanced Coulomb repulsion\cite{PhysRevB.28.117} in the high doping regime, generating the T$_c$ saturation or, in the most extreme case, a reentrant insulating state. This possibility is backed up by a number of observations: 

i) A progressively lower normal state carrier mobility  is measured in high doping regions in gated TMDs\cite{PMID:26235962,doi:10.1073/pnas.1716781115,Saito2016,Zheliuk2019}, indicating increasingly high disorder. Moreover, there is a clear correlation between the saturation of T$_c$ at high doping and a drop in the normal state mobility \cite{PMID:26235962,doi:10.1073/pnas.1716781115,Saito2016,Zheliuk2019}.

ii) The localization is expected to be stronger crossing from three to two dimensions. This is compatible with the weaker superconductivity (and much lower mobilities) measured in monolayer samples\cite{Costanzo2016,doi:10.1073/pnas.1716781115}. From this point of view the suppression of T$_c$ in monolayer samples would result from enhanced disorder and not from an intrinsic effect, in agreement with our theoretical calculations.

iii) The reentrant insulating phase in  WS$_2$ at high doping\cite{doi:10.1073/pnas.1716781115} resembles the metal-insulator transition of the Anderson model\cite{PhysRev.109.1492}.

Such a T$_c$ reduction with increasing resistivity has already been observed in A15 and other high-T$_c$ compounds\cite{PhysRevB.28.117}.
The enhancement in disorder results in a reduction of mobility and, thus, in a kinetic energy reduction. It has been suggested in Ref. \cite{PhysRevB.28.117}
that this effect could be modeled by using an increased Coulomb pseudopotential
(i.e. an increased Coulomb repulsion effectively reduces the electronic kinetic energies and mobilities).

We tested this hypothesis in Fig. \ref{fig25} where we looked at what values of the Coulomb pseudopotential $\mu^*$ are needed in single gated MoS$_2$ to reproduce the experimental T$_c$ versus doping curve in the saturation region (red points). We believe that this explanation is indeed possible.

 Another possible explanation comes from the observation that the predicted electronic structure as a function of doping for single-side gated MoS$_2$ is especially sensitive to lattice parameters. As such, the presence of (possibly doping dependent) in-plain strain or out-of-plane pressure in the experiments could reduce the critical temperature to some extent. 
 
 However, the universality of the observed dome trend, which is irrespective of the material, of the number of layers and of the substrate, suggests that a more general mechanism is at play, such as carriers localization.

\section{Conclusion}\label{sec7}
In this paper we presented a comprehensive first principles study of the superconducting properties of gated few-layer molybdenum dichalcogenides  as a function of the gate geometry (single or double gate) and of the sample (MoS$_2$ or MoSe$_2$). For the first time, all
relevant effects were included in the calculations, namely (i) the sample thickness, (ii) the field effect geometry, (iii) anharmonicity and (iv) the anysotropic and multband nature of the superconducting gap. This is a major advance with respect to previous works that neglected spin-orbit coupling, the effects of the electric field, of the sample thickness and anharmonicity.

Our work shows that the inclusion of the sample thickness, of the field effect geometry and of anharmonicity rules out the occurrence of charge density waves in the experimental doping range. The occurrence of CDW was indeed proposed as a mechanism for T$_c$ saturation. Furthermore,
we demonstrate a suppression of one order of magnitude 
(a factor $11.4$ at $n_e=1.38\times 10^{14}$ e$^-/{\rm cm}^{2}$ )
in the electron-phonon coupling with resopect to previous works \cite{Roesner2017}. Most important, the behaviour of our calculated electron-phonon coupling as a function of doping is in excellent agreement with transport data in Ref. \cite{AliElYumin2019}.

With this solid result at hand we explained the behaviour of Tc in different systems and
geometries. Interestingly we find that if only intrinsic effects were present, T$_c$ should increase
and not saturate in molybdenum dichalcogenides. This is an important result as if in the future experimental techniques allowing for a better control of the induced charge will be found, T$_c$ could raise even more.

Finally, by reviewing normal state mobility data and noting that the saturation of T$_c$ occurs always in region where there is a drop in mobility, we propose
that the Tc saturation is due to carriers localization and disorder that compete with phonon mediated superconductivity.

\section*{Acknowledgments}

We acknowledge support from the European Union's Horizon 2020 research and innovation programme Graphene Flagship under grant agreement No 881603. We acknowledge the CINECA award under the ISCRA initiative, for the availability of high performance computing resources and support. We acknowledge PRACE for awarding us access to Joliot-Curie at GENCI@CEA, France (project file number 2021240020). We thank Davide Romanin and Thibault Sohier for valuable advice regarding simulations in FET setup. MC acknowledges useful discussions with Ryosuke Akahashi.

\section*{Appendix: inadequacy of the uniform doping model} 

\begin{figure}[htpb!]
\includegraphics[width=0.9\linewidth]{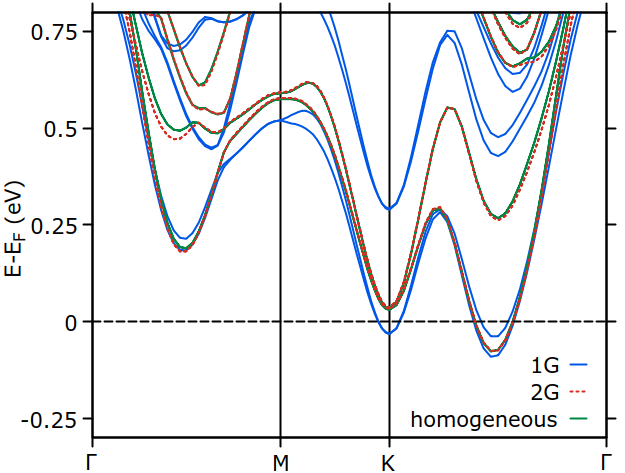}
    \caption{Comparison among the band structures for single-side gated MoS$_2$, double-side gated MoS$_2$ and homogeneously doped MoS$_2$ at $n_e=$1.44 $\times 10^{14}$ e$^-$/cm$^{2}$.}\label{fig26}
\end{figure}

In this section we demonstrate that the uniform doping (jellium) is completely inappropriate to describe single gate experiments, 
even for the case of a single layer in field-effect geometry.

To substantiate this statement we first compare the electronic structures of a MoS$_2$ bilayer at $n_e=$1.44 $\times 10^{14}$ e$^-$/cm$^{2}$ by using three setups: (i) uniform doping, (ii) single-side gated MoS$_2$ bilayer, and (iii) double-side gated MoS$_2$ . We observe that the electronic band structure obtained in the homogeneously doped case (green lines) is almost identical to the one obtained within the double-side gate setup (red dashed lines), while being completely different from the single-side gated one (blue lines). This demonstrates that homogeneously doped bilayer MoS$_2$ cannot be employed to obtain a meaningful comparison to single-side gating experiments. The agreement between the double gate setup and homogeneous doping suggests that homogeneous doping could be more appropriate in this case, at least for the electronic structure.

Even more striking is the fact that homogeneous doping does not reproduce well single-side gating even if a single-gate single-layer MoS$_2$ is considered. Surprising this is taken for granted in many publications, however this assumption is completely incorrect to describe vibration and superconducting properties.

As and example, the calculated electron-phonon coupling parameter $\lambda$ at $n_e=$1.15 $\times 10^{14}$ e$^-$/cm$^{2}$ is 1.23 by including explicitly the electric field, while it is $\lambda = 1.67$ in the homogeneously doped setup at the same doping. Thus, the charge inhomogeneity along the $z-$axis suppresses the electron-phonon coupling of $\approx 25\%$. 
This demonstrate the inadequacy of the homogeneous doping model.
\section*{References}
\bibliography{bibliography}
\bibliographystyle{iopart-num}

\end{document}